\documentclass[11pt,preprint]{aastex}
\usepackage{graphicx}
\usepackage{epsfig}
\usepackage{amsmath}
\usepackage{natbib}
\usepackage{url}
\usepackage{color,soul}

\slugcomment{Ap.J.\ in press}
\shorttitle{Signatures of Planets in Protostellar Disks}
\shortauthors{Isella A.\ and Turner N.}

\begin{document}

\title{Signatures of Young Planets\\ in the Continuum Emission From
  Protostellar Disks}
 
\author{Andrea Isella}
\affil{Department of Physics and Astronomy, Rice University\\ 6100
  Main Street, MS-108, Houston, Texas 77005}
\email{isella@rice.edu}
\and
\author{Neal J.\ Turner}
\affil{Jet Propulsion Laboratory, California Institute of
  Technology\\ 4800 Oak Grove Drive, Pasadena, California 91109}
\email{neal.turner@jpl.nasa.gov}

\begin{abstract}
Many protostellar disks show central cavities, rings, or spiral arms
likely caused by low-mass stellar or planetary companions, yet few
such features are conclusively tied to bodies embedded in the disks.
We note that even small features on the disk's surface cast shadows,
because the starlight grazes the surface.  We therefore focus on
accurately computing the disk's thickness, which depends on its
temperature.  We present models with temperatures set by the balance
between starlight heating and radiative cooling, that are also in
vertical hydrostatic equilibrium.  The planet has 20, 100, or
1000~M$_\oplus$, ranging from barely enough to perturb the disk
significantly, to clearing a deep tidal gap.  The hydrostatic balance
strikingly alters the model disk's appearance.  The planet-carved
gap's outer wall puffs up under starlight heating, throwing a shadow
across the disk beyond.  The shadow appears in scattered light as a
dark ring that could be mistaken for a gap opened by another more
distant planet.  The surface brightness contrast between outer wall
and shadow for the 1000-M$_\oplus$ planet is an order of magnitude
greater than a model neglecting the temperature disturbances.  The
shadow is so deep it largely hides the planet-launched spiral wave's
outer arm.  Temperature gradients are such that outer low-mass planets
undergoing orbital migration will converge within the shadow.
Furthermore the temperature perturbations affect the shape, size, and
contrast of features at millimeter and centimeter wavelengths.  Thus
radiative heating and cooling are key to the appearance of
protostellar disks with embedded planets.
\end{abstract}


\section{Introduction}
\label{sec:intro}

Extrasolar planetary systems are now known to be strikingly diverse in
the member planets' masses, multiplicities and orbital configurations
\citep{2007ARA&A..45..397U, 2013ApJS..204...24B, Laughlin2015,
  2015ARA&A..53..409W}.  Placing our Solar system in the context of
these new discoveries will require learning how the diversity arises
from differing conditions or processes in the protostellar disks that
gave the planets birth.  A promising way forward is by studying
spatially resolved observations of the protostellar disks orbiting
nearby young stars, which reveal numerous features in the emission
from the dust and gas: central cavities, bright and dark rings,
including asymmetric ones, and spiral arms.  These features are
qualitatively consistent with gravitational perturbation by low-mass
stellar or planetary companions, and offer an opportunity to learn
about the early stages of planet and star formation within the disks.

Clear examples of the variety of structures found in protostellar
disks are around the young stars SAO~206462 (also known as HD135344B),
J160421.7--213028 (hereafter J1604), and TW Hya.  The first two disks
have large ($>$10~AU) cavities that are dark in the continuum emission
from millimeter and centimeter-size grains.  While SAO~206462 shows
horseshoe-shaped azimuthal asymmetries \citep{Perez2014,
  2016ApJ...832..178V}, J1604 is characterized by an almost
axisymmetric narrow ring at millimeter wavelengths \citep{Zhang2014,
  2017ApJ...836..201D}.  In the optical and near-infrared, the
SAO~206462 disk shows a well-defined $m=2$ spiral design
\citep{Muto2012}, while the surface brightness around J1604 is
circularly symmetric \citep{2012ApJ...760L..26M, 2010ApJ...718L..87T},
similar in shape to the millimeter wavelengths.  Unlike SAO~206462 and
J1604, the TW~Hya disk at 0.85~mm shows a series of at least five
narrow concentric rings extending from 1 to 50~AU
\citep{2016ApJ...820L..40A}.  Near-infrared imaging also shows rings,
but these are much wider and only loosely related to those observed in
the sub-millimeter \citep{2017ApJ...837..132V}.
 
The dynamical interaction between the circumstellar material and
planets or low-mass stellar companions is the most widely discussed
explanation for the observed structures.  This is because, on one
hand, circular gaps, spiral waves, and azimuthal asymmetries are
naturally produced by the gravitational interaction with one or more
companions \citep{2000ApJ...533.1023L, 2014prpl.conf..667B,
  2012ARA&A..50..211K}, and on the other hand, the search for
exoplanets has revealed that most mature stars host one or more
planets, therefore many protostellar disks should show the effects of
planet formation.

Recent observations however point to major discrepancies with
planet-disk interaction models.  For example, the infrared spiral arms
observed in SAO~206462 and MWC~758 have large opening angles, which,
if interpreted as the manifestation of density waves propagating away
from an embedded planet at the local sound speed, imply temperatures
much higher than equilibrium with the stellar radiation field
\citep{Benisty2015}.  Furthermore, \citet{Juhasz2015} pointed out that
the appearance of the spiral structures in scattered light requires
disk pressure scale heights several times larger than expected under
hydrostatic equilibrium.  Finally, some models used to interpret the
observed spiral features lack internal consistency.  As an example,
planet masses of several Jupiters were derived in the case of
SAO~206462 by adopting prescriptions for the disk-planet interaction
that hold only for planets with masses less than 1~Earth.  In
contrast, the rather simple morphologies of the J1604 and LkCa~15
disks, and even the multi-ring morphology in TW Hya, HL~Tau,
HD~163296, and AS 209 can be explained using planet-disk interaction
models \citep{Dong2015a, 2015A&A...584A.110P, 2015A&A...584L...4P,
  2013ApJ...772...34J,2016ApJ...818...76J,Isella2016,2017arXiv171105185F}.
Furthermore, a candidate young planet has been imaged in the
dust-depleted cavity of LkCa~15 \citep{2015Natur.527..342S}.  Based on
this sample, we might argue that the features in ``dynamically cold''
ring-like disks result from planet-disk interaction, while those in
``dynamically hot'' asymmetric disks come from other yet unknown
physical processes.

Improving our understanding of planet-disk interaction requires both
enlarging the still small and biased sample of disks observed at high
angular resolution, and improving upon theoretical models of the
disks' response to perturbations from stellar and planetary
companions.  In this paper we address the modeling issue, motivated by
the fact that current treatments of embedded planets' effects on the
disk's appearance involve very simple prescriptions for the disk's
thermal response.  \citet{Juhasz2015} noted that spiral structures in
the scattered light are more sensitive to disturbances in the disk's
pressure scale height than to disturbances in the density.  The scale
height is proportional to the sound speed and thus depends on the
temperature.  The temperature in turn is highly sensitive to the angle
of incidence of the starlight, so accurately calculating the shape of
the disk surface is crucial for modeling the system's appearance
\citep{2008ApJ...679..797J}.

To address these issues, we investigate the planet-disk interaction's
effects on the disk temperature and density.  We construct models that
are in hydrostatic equilibrium in the vertical direction, with
temperatures set by the balance between starlight heating and
radiative cooling.  Two-dimensional hydrodynamical calculations in the
equatorial plane yield surface density maps, which we expand into
three-dimensional density distributions using the hydrostatic
equilibrium for the initially guessed temperature distribution.  We
compute new temperatures using detailed frequency-dependent Monte
Carlo radiative transfer calculations treating scattering, absorption
and thermal re-emission.  We then step forward in time, letting each
patch of the disk expand or contract vertically on its own thermal or
dynamical timescale, and again compute temperatures accounting for the
way the starlight falls across the disk's new shape.  The calculations
continue as long as needed to reach joint radiative and hydrostatic
equilibrium.  This modeling approach is laid out in section~2, and we
describe its results in section~3.  We find that planets affect the
temperatures through the interplay between starlight heating, the
disk's shape, and radiative cooling.  We then apply the model to
investigate the planet-induced perturbations' observational
appearance.  We focus on signatures in infrared scattered light, which
trace the disk's surface layers, and in millimeter-wave dust continuum
emission, which trace the dust in the midplane.  Section~4 deals with
the temperature structures' implications for disks' equation of state
and planets' orbital migration.  The models' relationship to the
features observed in protostellar disks is discussed in section~5, and
the conclusions are presented in section~6.

\section{Modeling the Planet-Disk Interaction}

Accurately modeling the planet-disk interaction requires solving the
following problem.  The orbiting planet exerts a gravitational force
on nearby disk material, launching acoustic waves which compress and
rarefy the gas through which they pass, perturbing the surface
density.  The density fluctuations alter the height and slope of the
surface where the illuminating starlight is absorbed.  Large enough
excursions of the surface cast shadows across the disk beyond.  The
changing illumination is reflected in the system's appearance in
scattered light.  Over the heating or cooling timescale, the changing
illumination also affects the interior temperatures, which feed back
on the acoustic waves' own propagation.  Furthermore, planets more
massive than Jupiter launch waves strong enough to clear disk material
from an annulus around the planetary orbit.  This drastic change in
the disk structure further alters how starlight is absorbed and
scattered across the disk.

Several approximations are necessary to make this problem tractable.
To begin with, we assume that the planet disturbs the disk in a way
that depends weakly on the temperature, so we can solve for the gas
dynamics using temperatures differing from those we will eventually
determine through detailed Monte Carlo radiative transfer.  We find
that this is valid for planets too low in mass to open a gap in the
disk.  However the formation of a gap leads to large temperature
excursions, making the approximation questionable.  We nevertheless
separate the dynamics and radiative transfer because the latter takes
far too much computer time to be carried out with each hydrodynamical
time step.  Frequency-averaged semi-analytic transfer methods are much
faster \citep{2009ApJ...700..820J} and adequately reproduce the
results of detailed transfer calculations \citep{2012ApJ...749..153J},
but have not been combined with the 3-D hydrodynamical methods needed
to model the surface-illuminated disk's non-axisymmetric response to
an embedded planet.  Numerical 3-D radiation hydrodynamics modeling is
now feasible in the flux-limited diffusion approximation, including
direct starlight treated by integrating along radial rays
\citep{2010A&A...511A..81K, 2011A&A...536A..77B, 2013A&A...560A..43F},
but the flux-limiting yields wrong answers when shadowing is important
\citep{2003ApJS..147..197H}, and these approaches also neglect
scattered starlight.  In contrast, the Monte Carlo approach finds the
radiation field's full dependence on both angle and frequency.  Thus
the models for gap-opening planets presented below, despite their
limitations, provide insights into the appearance of disks with
embedded planets that are not available from other existing results.

\subsection{Unperturbed Disk}
\label{sec:unperturbed_model}

We start our investigation by adopting a disk model unperturbed by any
planet, with gas surface density
\begin{equation}
  \Sigma_g^0(r) = 18\ \textrm{g cm}^{-2} \times
  \left(\frac{r}{10 \textrm{AU}}\right)^{-1}
  \label{eq:sigma}
\end{equation}
extending from 0.1~AU to 50~AU.  Its total mass is 0.007~$M_\odot$,
less than the minimum-mass Solar nebula but consistent with the masses
of young disks in nearby star forming regions
\citep[e.g.][]{Andrews07,Isella09}.

The central star has a mass of 1~$M_\sun$, a luminosity of 1~$L_\sun$,
and an effective temperature of 5600~K.  The disk temperature and
vertical density structure are calculated under the assumption of
hydrostatic equilibrium between the gas pressure and the stellar
gravity.  Dust is well-mixed throughout the gas, with a uniform
gas-to-dust ratio of~100.  The interaction between dust and gas is
further discussed in Section~\ref{sec:dust-gas}.  Temperatures are
computed using the Monte Carlo radiative transfer code RADMC-3D.  The
disk is placed in radiative equilibrium, so that the bolometric
emission from each grid cell matches the rate at which starlight and
other incident radiation are absorbed, following the procedure
discussed in Section~\ref{sec:disk_temp_mod}.

The midplane temperature of the resulting unperturbed disk model is
well approximated by the power-law fit
\begin{equation}
  T_m^0(r) \simeq \textrm{41}\textrm{K} \times \left(
  \frac{r}{10\textrm{AU}}\right)^{-0.45}.
  \label{eq:t_0}
\end{equation}
The temperature along the direction perpendicular to the midplane is
equal to $T_m^0(r)$ for $z/r < h/r \approx 0.04$.  At greater heights
the temperature increases to reach that of a dust grain in optically
thin surroundings exposed to the starlight, $T_s^0(r) \simeq 124
\,\textrm{K}(r/10\textrm{AU})^{-1/2}$ \citep[e.g.][]{Dullemond01,
  Dullemond04}.

Since most of the disk's mass lies in the vertically-isothermal
interior, the vertical density profile to first approximation is the
Gaussian
\begin{equation} 
  \rho(r,z) = \frac{\Sigma(r)}{\sqrt{2\pi}h(r)} \exp\frac{-z^2}{2h^2(r)}.
  \label{eq:rho}
\end{equation}
At hydrostatic equilibrium, the pressure scale height $h(r)$ is equal
to $c_s(r)/\Omega(r)$, where $c_s(r)=\sqrt{k_b T(r)/\mu m_p}$ is the
isothermal gas sound speed, and $\Omega(r)=\sqrt{GM_\star/r^3}$ is the
Keplerian angular speed.  Using the gas temperature from
Equation~\ref{eq:t_0} and a mean molecular weight $\mu=2.3$, the sound
speed and pressure scale height are
\begin{equation}
  c^0_s(r) \simeq 0.38\ \textrm{km s}^{-1} \times
  \left(\frac{r}{10\textrm{AU}}\right)^{-0.225}
  \label{eq:cs}
\end{equation}
and
\begin{equation}
  h^0(r) \simeq 0.41 \textrm{AU} \times
  \left(\frac{r}{10\textrm{AU}}\right)^{1.275}.
  \label{eq:h_p}
\end{equation}

Below we use the fits in Equations~\ref{eq:sigma} and \ref{eq:t_0} as
baselines against which to compare the structure of model disks
perturbed by planets.
\\*[1mm]

\subsection{Planets Perturbing the Disk}
\label{sec:p-d}
The first step in producing synthetic images of a disk interacting
with an embedded planet is to map the perturbations in the surface
density.  If the planet is low-mass, the perturbations are small and
can be calculated by linearizing the equations of motion and
continuity \citep{1978ApJ...222..850G}.  In contrast, high-mass
planets induce non-linear perturbations that must be investigated
using numerical simulations \citep{Bryden99}.  The transition between
these two regimes occurs when non-linear effects, such as shocks,
become relevant in transferring the angular momentum carried by the
waves into the gas \citep{2001ApJ...552..793G}.

The transition from linear to non-linear perturbations in a Keplerian
disk occurs around the thermal mass $M_{th} = 2c_s^3/3\Omega G$
\citep{2001ApJ...552..793G, 2011ApJ...741...57D}.  A planet with mass
$M_p = M_{th}$ has a Hill sphere whose radius is comparable to the
disk's pressure scale height.  In an inviscid disk, $M_{th}$ is the
mass at which planets start to open annular gaps in the surrounding
gas by depositing angular momentum at their innermost Lindblad
resonance \citep{1993prpl.conf..749L}.

For the unperturbed disk model described above, the critical mass is
\begin{equation}
  M_{th} \sim 15 M_\oplus \times
  \left(\frac{r}{\textrm{10AU}}\right)^{0.825}.
\end{equation}
If $M_p \ll M_{th}$, the planet perturbation's waveform can be written
in polar coordinates as
\begin{equation}
  \phi = \phi_p +
  \textrm{sign}(r-r_p)\frac{r_p}{h_p} \times
  4 \left[ \left(\frac{r}{r_p}\right)^{-0.275} 
    + \frac{1}{5} \left(\frac{r}{r_p}\right)^{1.225}
    - 1.48 \right],
  \label{eq:raf}
\end{equation}
where $\phi_0$ is the planet's position angle and $h_p$ is the
pressure scale height at the planet's orbital radius $r_p$.  This
equation was derived from \citet{2002ApJ...569..997R} Equation~44
using a power law index for the sound speed $\nu=0.225$, as in
Equation~\ref{eq:cs}.  In the linear regime, the perturbation is a
one-armed spiral propagating from the planet both inward and outward
and characterized by an opening angle depending only on the sound
speed.  The density perturbation's maximum amplitude $\delta\Sigma /
\Sigma_0 \simeq M_p/M_{th} \ll 1$, where $\Sigma_0$ is the unperturbed
surface density profile.

If $M_p\geq M_{th}$, the density perturbations induced by the planet
must be calculated by numerically solving the hydrodynamics equations.
In this paper, we employ the publicly available code FARGO3D
\citep[][\url{http://fargo.in2p3.fr}]{Masset2000} to simulate the
disk's interaction with planets having masses of 20, 100 and
1000~$M_\oplus$, corresponding to about 1.3, 7 and 70~times the thermal mass
respectively.  Henceforth, we refer to these three hydrodynamical
results as P20, P100, and P1000, respectively.  We arbitrarily place
the planets at 10~AU so that the perturbations on the disk structure
have spatial scales observable with current infrared and
millimeter-wave telescopes.  The calculations are two-dimensional in
the orbital plane, with the disk structure averaged over the
perpendicular direction.  The code setup is described in
appendix~\ref{appendix:fargo}.

\begin{figure*}[!t]
\centering
\includegraphics[angle=0, width=\linewidth]{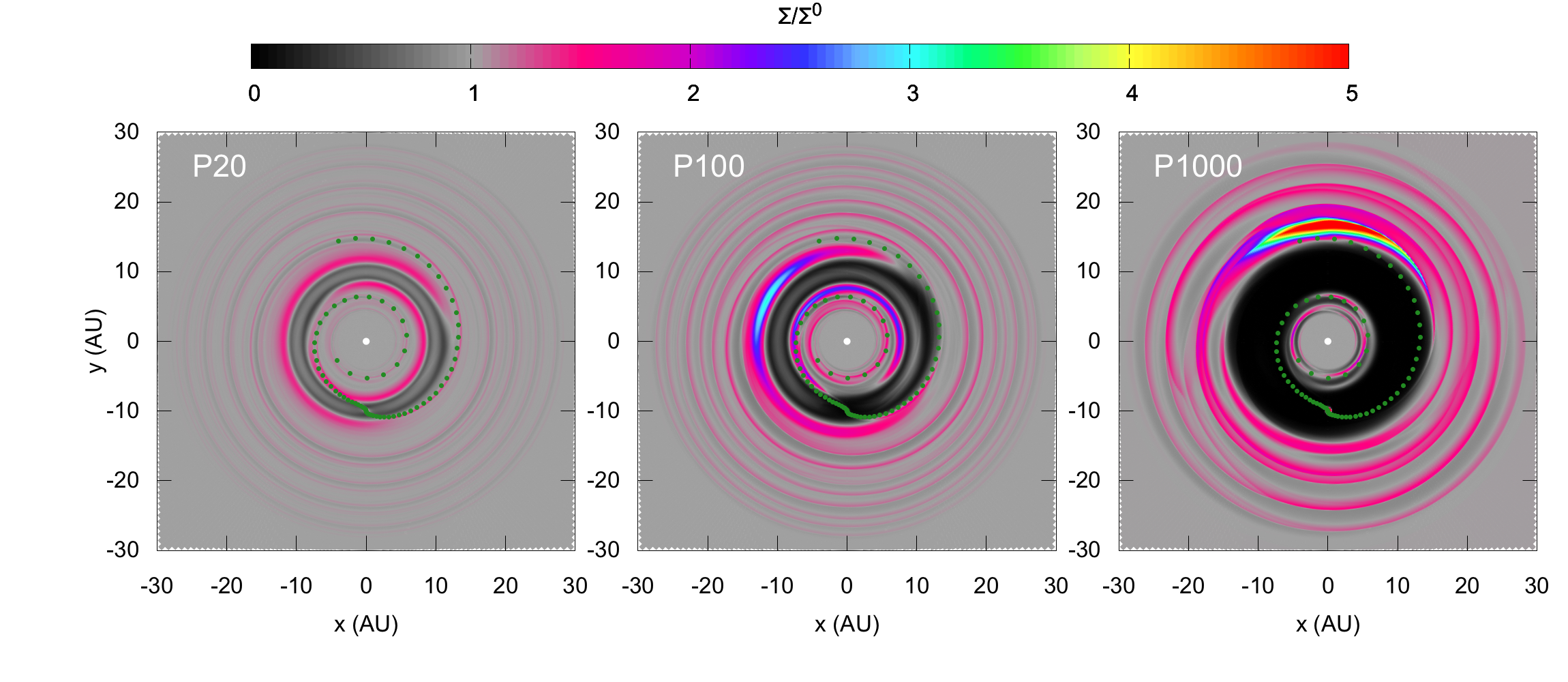}
\caption{\label{fig:fargo} Surface density perturbations induced by
  planets of 20 (P20), 100 (P100) and 1000 (P1000) M$_\oplus$ (left,
  center, and right panels, respectively) orbiting 10~AU from the
  star, after about 300~planet orbits ($\sim 10^4$~yr).  The color
  scale shows the amplitude of the perturbations $\Sigma / \Sigma^0$,
  relative to initial surface density profile $\Sigma^0$ defined by
  Equation~\ref{eq:sigma}.  Green dots show the locus of the peak of
  the spiral density wave from the linear theory
  (Equation~\ref{eq:raf}). }
\end{figure*}

In Figure~\ref{fig:fargo} are snapshots of the surface density in the
P20, P100, and P1000 models after 300~orbits of the planet.  To
highlight the perturbations, we plot the ratio of the gas surface
density to that of the unperturbed disk defined by
Equation~\ref{eq:sigma}.  In the P20 model ($M_p \sim M_{th}$), the
planet creates an $m=1$ spiral wave, similar in shape and amplitude to
the linear solution of Equation~\ref{eq:raf}.  Furthermore, the planet
partially depletes a circular gap, reducing the surface density to
about half its initial value over an annulus about 3~AU wide.  Near
the gap's inner and outer edges are horseshoe structures in which the
surface density varies by about 40\% with azimuth.

More massive planets lead to stronger density perturbations.  The gap
is wider and more highly depleted, the spiral density wave has a
higher amplitude and deviates more from the linear solution, and the
azimuthal asymmetries grow.  The non-axisymmetric structures take the
form of anticyclonic vortices \citep{Li00}.  It is worth noting that
while spiral density waves rotate at the planet's Keplerian speed, the
vortices formed at the gap edges rotate at the local orbital speed.
This difference could profoundly impact the coupling between dust and
gas, the disk thermal structure, and therefore the observability of
such perturbations.

Finally, we point out that our hydrodynamic simulations' outcomes
depend on poorly known physical quantities such as the accretion
stresses, as well as numerical parameters such as the smoothing length
applied to the planet's gravitational potential (see
appendix~\ref{appendix:fargo}).  Investigating the interplay of these
parameters is not the purpose of this paper.  Instead, we use FARGO3D
simply as a tool to produce surface density maps showing features
similar to those observed in protostellar disks, namely, circular
gaps, azimuthal asymmetries, and spiral density waves.

\subsection{Dynamical Coupling Between Dust and Gas}
\label{sec:dust-gas}

While the planet-disk interaction affects the distribution of
circumstellar gas, the bulk of the disk's opacity is carried by dust
grains.  The perturbations that planets induce on the disk temperature
and continuum emission therefore depend on the relative distributions
of dust and gas.  The two are coupled by drag forces over the stopping
time $\tau_s$, which in the case of a Keplerian disk in hydrostatic
equilibrium, can be written as
\begin{equation}
  \label{eq:ts}
  \tau_s
  \approx {2\rho_ia\over\Sigma\Omega}
  \sim 1.1 \textrm{yr} \times \left(\frac{a}{1\mathrm{cm}}\right)
  \left(\frac{r}{10\textrm{AU}}\right)^{5/2}
\end{equation}
\citep[e.g., the review by][]{2014prpl.conf..339T}, where $a$ is the
particle size and $\rho_i$ is the grains' internal density.  The
right-hand form of the equation results from adopting a grain density
of 2~g~cm$^{-3}$ and the surface density profile in
Equation~\ref{eq:sigma}.  The stopping time increases linearly with
the grain size, so small particles are more closely coupled to the gas
than big ones.  To quantify the effect of planet perturbations on
particles of different sizes, we compare the stopping time $\tau_s$ to
the characteristic time scale for density perturbations $\tau_e$
calculated as follows.  Spiral waves excited by a planet rotate at the
Keplerian speed $\Omega_p = \sqrt{GM_\star/r_p^3}$.  The relative
angular speed between the spiral perturbation and a grain orbiting at
a distance $r$ from the star is therefore
\begin{equation}
  \delta\Omega(r)
  = \Omega(r) - \Omega_p
  = \sqrt{GM_{\star}} \left(\frac{1}{r^{3/2}}-\frac{1}{r_p^{3/2}}\right).
\end{equation}
If the density perturbation has a characteristic angular width
$\delta\phi$, then during each orbit, a dust grain is entrained in the
density wave only for the characteristic time
\begin{equation}
  \label{eq:te}
  \tau_e
  \simeq \frac{\delta\phi}{\delta\Omega}
  = \frac{\delta\phi}{\sqrt{GM_{\star}}}
  \left|\frac{1}{r^{3/2}} - \frac{1}{r_p^{3/2}}\right|^{-1}.
\end{equation}

\begin{figure*}[!t]
\centering
\includegraphics[width=0.7\textwidth]{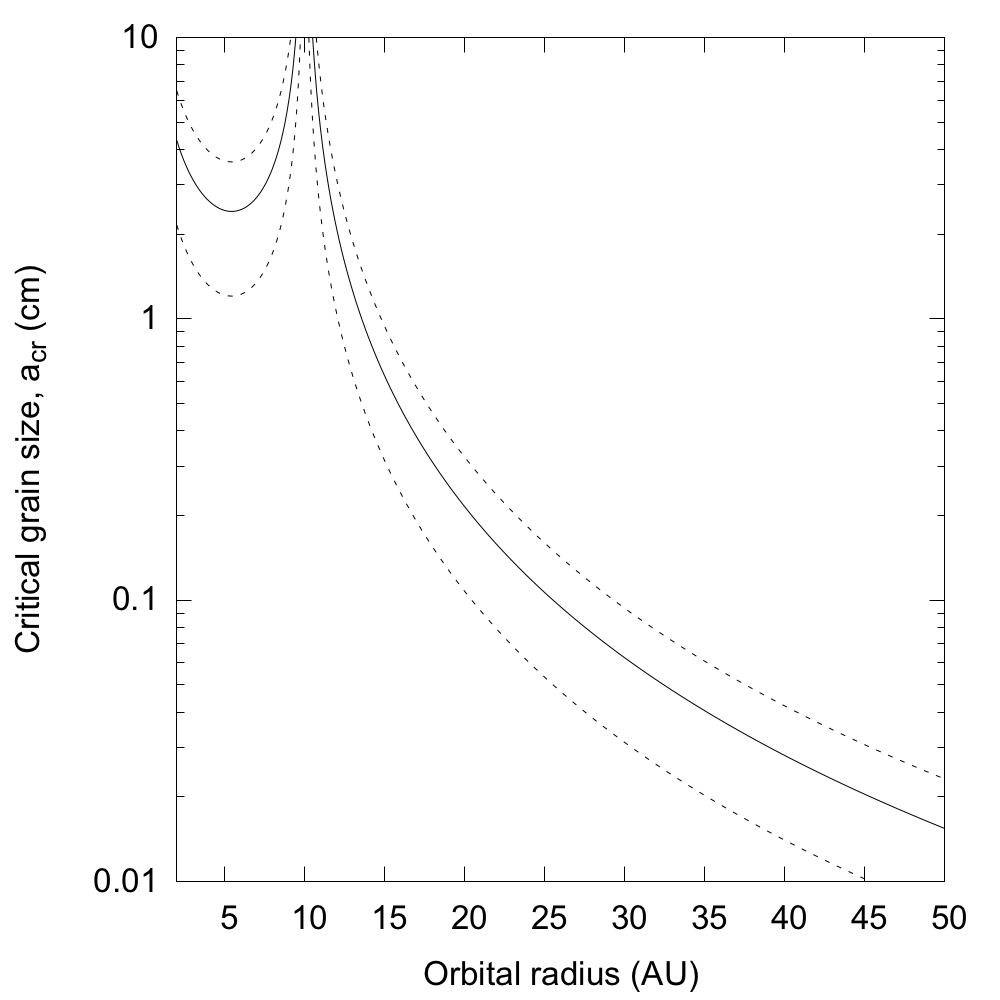}
\caption{Critical dust particle size vs.\ distance from the star, for
  a planet orbiting at 10~AU.  Only particles smaller than $a_{cr}$
  couple to spiral density perturbations raised by the planet.  The
  solid line corresponds to an angular extent for the perturbation
  $\delta\phi=10^\circ$, and the upper and lower dashed lines to
  $\delta\phi=15^\circ$ and $\delta\phi=5^\circ$, respectively.}
\label{fig:acr}
\end{figure*}

Particles with $\tau_s\ll\tau_e$ will be distributed in the same way
as the gas.  If the gas surface density increases by $\delta\Sigma$,
the dust surface density will simultaneously increase by the same
proportion.  Particles with $\tau_s\gg\tau_e$ on the other hand will
not respond to the gas density variations.  In between is the critical
grain size $a_{cr}$, at which $\tau_s=\tau_e$.  Particles smaller than
$a_{cr}$ trace the gas disturbances, while larger particle are free to
respond to forces other than the gas drag.  Using Equation~\ref{eq:ts}
and \ref{eq:te}, we write the critical grain size as
\begin{equation}
  a_{cr}(r) =
  \delta\phi\times\frac{\Sigma(r)}{2 \rho_i} \times
  \left|1-\left(\frac{r}{r_p}\right)^{3/2}\right|^{-1} \simeq
  0.8\ \textrm{cm} \times \left(\frac{10{\textrm{AU}}}{r}\right)
  \times \left|1-\left(\frac{r}{r_p}\right)^{3/2}\right|^{-1},
\end{equation}
where in writing the rightmost form, we assumed $\delta\phi =
10$\arcdeg.  Figure~\ref{fig:acr} shows $a_{cr}$ for $r_p=10$~AU and
$\delta\phi=5$, 10 and 15\arcdeg.  The critical grain size at 20~AU
and 40~AU from the star ranges from 0.1 to 0.3 and 0.01 to 0.03~cm,
respectively.  This suggests that the grains dominating the cm- and
mm-wave continuum emission will trace spiral density waves generated
by a planet only close to the planet itself.  Micron-sized particles,
which are much smaller than the critical size, are well coupled to
these perturbations throughout the disk.

The formation of gas-depleted gaps and vortices also affects the
distribution of the dust relative to the gas.  A gap's outer edge is a
local maximum in the gas pressure, and thus traps dust particles
\citep{2003ApJ...583..996H}.  Trapping is most efficient for particles
with Stokes number $St = \tau_s\Omega \simeq 2\rho_ia/\Sigma$ close to
unity.  With the surface density profile in Equation~1, $St = 1$
corresponds to a grain size $a(St=1) = 4.5\ \textrm{cm} \times (r/10
\textrm{AU})^{-1}$.  A gap's outer edge will therefore preferentially
trap cm- and mm-sized particles, while micron-sized dust remains
coupled to the gas.  Similarly, the gas pressure maxima at the centers
of the vortices that develop on the gap's outer edge preferentially
concentrate large dust particles \citep{2009A&A...497..869L,
  Birnstiel13}.

In summary, dust grains smaller than a few microns are likely to be
well coupled to the gas through drag forces even in the presence of
time-variable perturbations like those created by a planet.  These
small grains carry most of the opacity at optical and infrared
wavelengths, and therefore control the starlight absorption and the
disk's temperature.  However the continuum emission at mm and cm
wavelengths comes from dust particles that are big enough to decouple
from the gas.  Thus, combining spatially-resolved observations of the
gas and dust continuum emission at near-infrared and mm to cm
wavelengths can potentially reveal both planet signatures and the
dynamics of the larger particles.

\subsection{Calculating Temperatures in Disks Perturbed by Planets}
\label{sec:disk_temp_mod}

In this section, we describe the approach we adopt to compute the
temperatures in disks perturbed by planets.  We start by assuming that
the disk beyond a few AU is heated only by the starlight.  Accretion
heating is neglected, which is a good approximation outside 1~AU for
mass accretion rates less than about $10^{-8}$~M$_\odot$ yr$^{-1}$
\citep{Dalessio98}.  We also assume that shock and compression heating
related to the planet perturbations are negligible compared to the
starlight heating.  In this limit, the temperatures depend on the
vertical distribution of dust particles smaller than a few $\mu$m,
which dominate the opacity at optical and infrared wavelengths.
Following the discussion of the previous section, we assume these
particles are dynamically well-coupled to the gas.  We take the dust
and gas temperatures to be equal, which is a good approximation at
vertical visual extinctions $A_V$ greater than about 0.1~magnitude
\citep{2007prpl.conf..555D}.

For smooth surface density profiles like that of
Equation~\ref{eq:sigma}, the balance between gas pressure and stellar
gravity leads to a hydrostatic equilibrium solution in which the
disk's top and bottom surfaces are concave, bending upward with
increasing distance from the star.  The temperature and the pressure
scale height follow radial power-laws similar to
Equations~\ref{eq:t_0}-\ref{eq:h_p} \citep{Chiang97, Dalessio98,
  Dullemond01}.  By contrast, disks with non-monotonic surface density
profiles like those resulting from disk-planet interactions have had
their temperatures calculated only in a few cases.
\citet{2012ApJ...748...92T} examined Jupiter's effects on the Solar
Nebula, finding that the gap opened by the planet was up to twice as
hot as the same location in an unperturbed Solar nebula.  The low
optical depth lets starlight scattered and reprocessed on the gap's
walls reach the midplane.  Such planet-driven temperature
perturbations could impact the chemistry of the circumplanetary
material, and perhaps the evolution of the disk itself.

Building on this result, we develop a numerical scheme to calculate
the three-dimensional thermal structure of a disk perturbed by one or
more planets.  Our method is based on the consideration that the
disk's response to changes in illumination depends on the ratio of the
thermal and dynamical time scales.  The thermal time scale $t_{th}$ is
the time for the disk to heat or cool to a new thermal equilibrium,
and the dynamical time scale $t_{dy}$ is the time for the disk to
reach hydrostatic equilibrium.  Following \citet{2008ApJ...672.1183W},
we consider the region beyond about 1~AU from the star where accretion
heating can be neglected, and write the thermal and dynamical time
scales as
\begin{equation}
  t_{th} = \frac{(\gamma+1)}{2(\gamma-1)}
  \frac{c^2_s \Sigma}{\sigma T^4}
  \propto r^{0.35}
\end{equation}
and
\begin{equation}
  t_{dy} = \frac{2\pi}{\Omega} \propto r^{1.5},
\end{equation}
where $\gamma$ is the adiabatic index of the gas. 

\begin{figure}[!t]
\centering
\includegraphics[angle=0, width=0.7\linewidth]{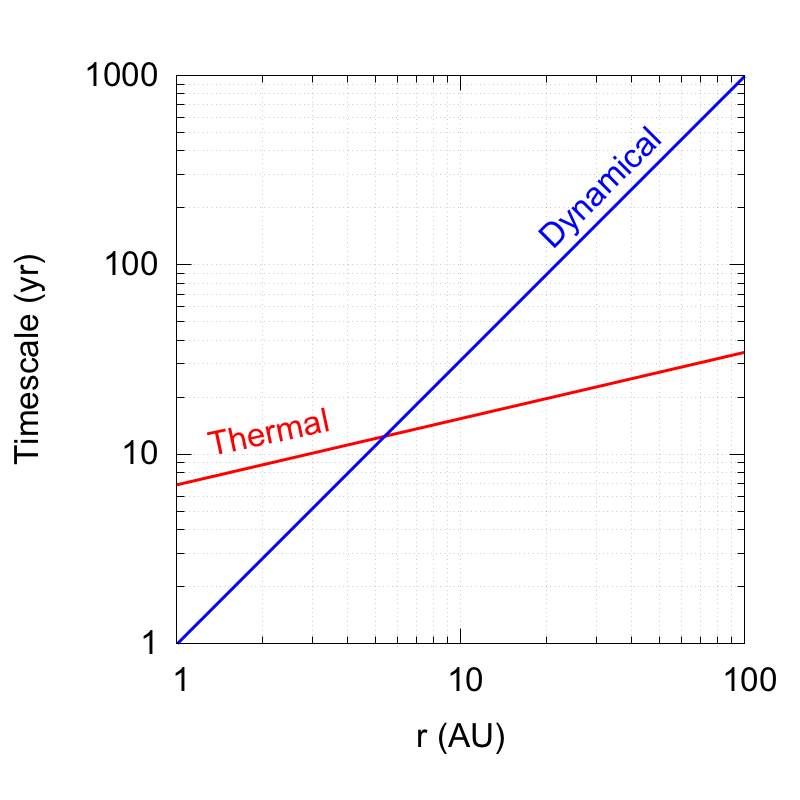}
\caption{Thermal $t_{th}$ and dynamical $t_{dy}$ timescales
  vs.\ radius in the unperturbed disk model described in
  Section~\ref{sec:unperturbed_model}.}
\label{fig:timescale}
\end{figure}

Figure~\ref{fig:timescale} shows the ratio of thermal to dynamical
time for the unperturbed disk model described in
Section~\ref{sec:unperturbed_model}.  The two time scales reach the
same value of about 13~yr at $r\simeq 5.5$~AU.  We call these the
crossover time scale $t_{cr}$ and radius $r_{cr}$, respectively.  At
$r\ll r_{cr}$, where $t_{dy}\ll t_{th}$, the disk can expand or
contract in the vertical direction faster than its temperature can
change.  Thus after each change in the illumination, the disk will
gradually approach thermal equilibrium while remaining close to
hydrostatic equilibrium throughout.  In the opposite extreme, at $r\gg
r_{cr}$ where $t_{dy}\gg t_{th}$, the vertical structure varies slower
than the temperature, so under changing illumination, the disk will
quickly reach thermal equilibrium while spending some time out of
hydrostatic equilibrium.
  
The following example illustrates these two time scales' effects.
Imagine that a planet orbiting near the crossover radius opens a gap,
creating a local maximum in the gas and dust density near the gap's
inner edge.  The higher dust density means more starlight is
intercepted here, so the disk temperature begins increasing.  Since
the dynamics are quicker than the heating, the disk promptly expands
to the hydrostatic equilibrium corresponding to the rising
temperature.  As its height increases, the gap's inner edge casts a
taller shadow on the disk beyond.  Under the decreased illumination,
the outer regions (i.e.\ the regions outside the crossover radius)
quickly cool off.  The shadowed parts contract toward the midplane on
the local dynamical time scale, which is longer than the local thermal
timescale, so even after the gap's inner edge reaches thermal and
hydrostatic equilibrium, the outer disk is still contracting towards
the thinner state appropriate for its new cooler temperature.

To track the disk's response to the starlight across both regimes ---
with the thermal time scale the slower of the two, and with the
dynamical time scale the limiting factor --- we implement a sort of
poor man's radiation hydrodynamics.  First, we bring the 2-D
vertically-averaged hydrodynamics calculation with embedded planet to
an approximate steady-state, and construct an initial guess at the 3-D
density structure using the scale height $h^0(r)$ from the same radius
in the unperturbed disk (Equation~\ref{eq:h_p}).  Second, we send
starlight into this structure using Monte Carlo radiative transfer, to
determine new temperatures.  Third, we adjust the structure of each
patch of disk towards vertical hydrostatic equilibrium, letting the
scale height change no more than permitted by the ratio of the time
step to the local limiting time scale.  Then we continue stepping
forward in time.  With each time step we compute first the new
temperatures by Monte Carlo transfer, and then the new densities by
letting the gas expand or contract towards hydrostatic balance.  This
simplified dynamics lets us determine whether the outer disk settles
into a solution where it is (1) starlit, warm, and flared, or (2)
shadowed, cold, and flat \citep{Dullemond04}.  The final solution is
both in radiative balance with the starlight, and in vertical
hydrostatic equilibrium.  Note that the disk can reach such a joint
equilibrium only if every patch is illuminated steadily for longer
than both its dynamical and thermal timescales.  Faster-varying
lighting, for example due to shadows cast by non-axisymmetric
structures orbiting nearer the star, can maintain permanent
disequilibrium, but this is not captured here.

\begin{figure*}[!t]
\centering
\includegraphics[angle=0, width=0.49\linewidth]{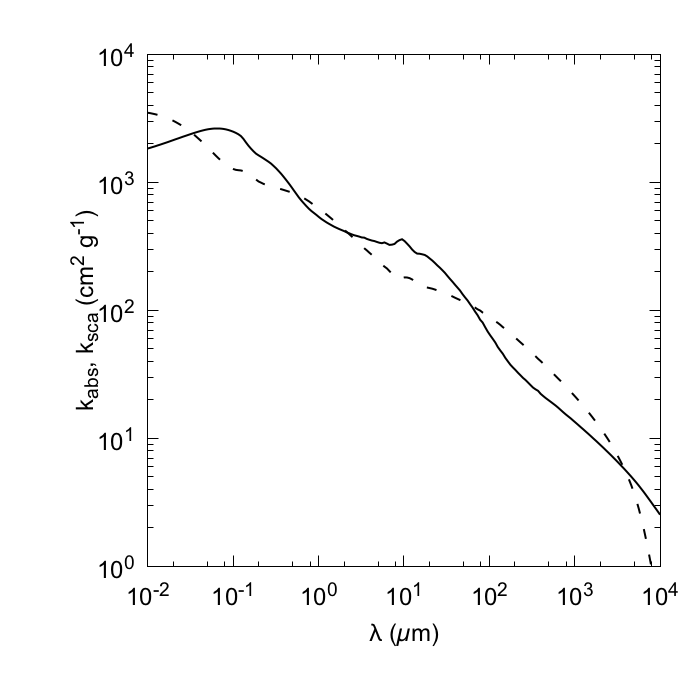}
\includegraphics[angle=0, width=0.50\linewidth]{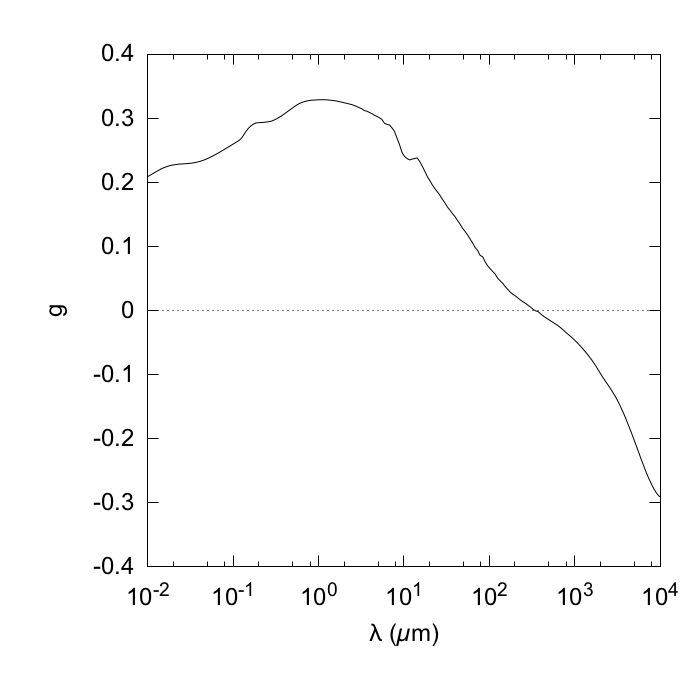}
\caption{Left: Wavelength-dependent absorption (solid) and scattering
  (dashed) opacities adopted in calculating the disk's temperature and
  appearance.  Right: Wavelength-dependent scattering asymmetry
  parameter $g$.}
\label{fig:dust_opac}
\end{figure*}

For the radiative transfer piece of each time step, we use the Monte
Carlo code RADMC-3D.  This follows the paths of large numbers of
photons as they are absorbed and scattered by the dust.  The code is
set up as spelled out in appendix~\ref{appendix:radmc}.  Heating and
cooling are calculated by adopting absorption and scattering opacities
for spherical grains made of a mixture of astronomical silicates and
carbonaceous materials, with relative abundances as in
\citet{Pollack94}.  The materials' optical properties are mixed
together using the Bruggeman theory to calculate the opacities for
grains of a single size, which are then averaged over the size
distribution $n(a)\propto a^{-3.5}$, with $a$ ranging from 0.01 to
1000~$\mu$m.  Anisotropic scattering is treated using the
Henyey-Greenstein phase function.  The absorption and scattering
opacities, and asymmetry parameter $g$, are calculated using the Mie
theory and are shown in Figure~\ref{fig:dust_opac}.  We have varied
the dust composition and the grain size limits, with little effect on
the disk's final temperature structure.

For the gas dynamics piece of the time step, we evolve the density
profile of each disk patch toward vertical hydrostatic equilibrium.
We integrate the force balance equation from the midplane upwards, but
let the density scale height at each point change by a fraction no
greater than the time step divided by either the thermal or dynamical
timescale, whichever limits the gas movements.  The new density
profile is closer to, but not necessarily in, hydrostatic equilibrium.
To conserve mass, we adjust the midplane density at the start of the
integration till the new structure has the same surface density as the
old.  The method is spelled out in full in
appendix~\ref{appendix:pmhd}.

Compared with \citet{2011A&A...536A..77B} and
\citet{2013A&A...560A..43F}, we thus treat the gas flows in a highly
simplified fashion, while giving the radiation field the comprehensive
Monte Carlo transfer treatment.  The method yields three kinds of
model disks which we examine below: (1) in radiative equilibrium with
the starlight, and in vertical hydrostatic balance, but unperturbed
and planet-free, (2) disturbed by an embedded planet, and placed in
radiative equilibrium with the starlight, but having the same scale
heights $h^0(r)$ as the unperturbed model, so that it is
hydrostatically out of balance, and (3) with the embedded planet, and
placed in joint radiative and hydrostatic equilibrium using the poor
man's radiation hydrodynamics approach.  We refer to these as models
RH, P20R, and P20RH, respectively, when the planet has 20~M$_\oplus$.
Thus our complete procedure yields a total of seven models.  The other
four are P100R, P100RH, P1000R, and P1000RH.

\section{Results}

\subsection{Planet Perturbations' Effects on Disk Temperatures}
\label{sec:temp}

\begin{figure*}[!t]
\centering
\includegraphics[width=1.0\linewidth]{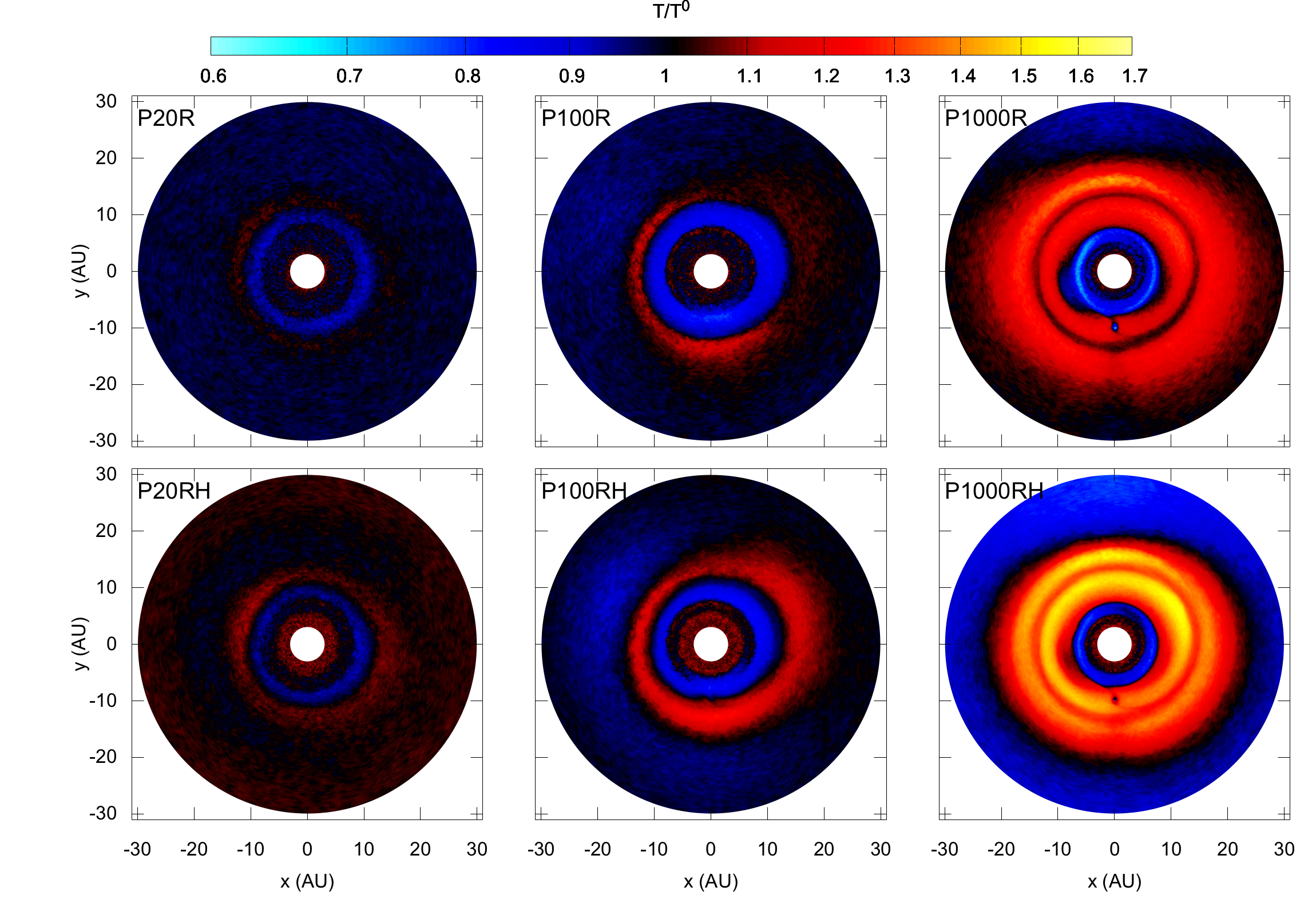}
\caption{Midplane temperature perturbations in the models with planets
  of 20 (left), 100 (center), and 1000~M$_\oplus$ (right).  The top
  row shows models P20R, P100R and P1000R with the pressure scale
  height unchanged from the planet-free disk.  The bottom row shows
  models P20RH, P100RH and P1000RH with pressure scale heights found
  by placing the disk in hydrostatic as well as radiative
  equilibrium.}
\label{fig:temp_radmc}
\end{figure*}

\begin{figure*}[!t]
\centering
\includegraphics[angle=0, width=1\linewidth ]{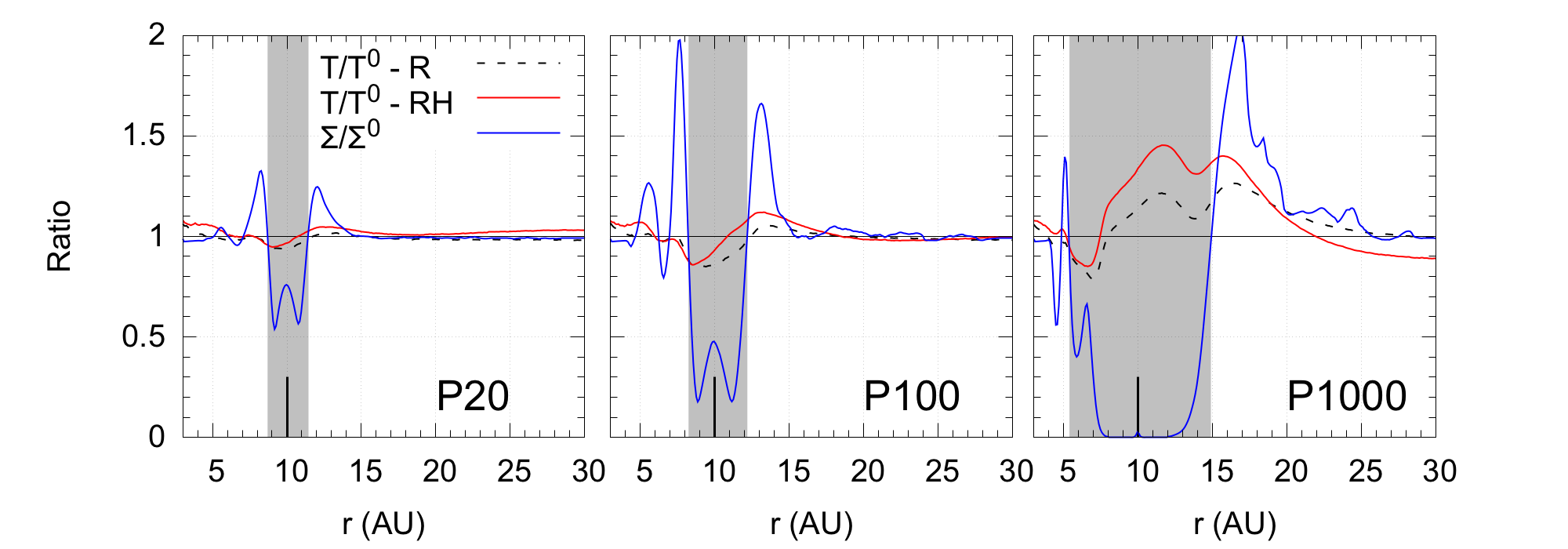}
\caption{Azimuthally-averaged surface density and midplane temperature
   profiles in the 20-, 100-, and 1000-M$_\oplus$ cases (left to
   right), divided by the planet-free model RH profiles.  Dashed lines
   show models P20R, P100R and P1000R with the unperturbed scale
   heights, and solid lines are models P20RH, P100RH and P1000RH in
   hydrostatic as well as radiative equilibrium.  Gray shading marks
   the gap opened by the planet, where surface densities are less than
   in the planet-free case.}
\label{fig:temp_2d}
\end{figure*}

Figure~\ref{fig:temp_radmc} shows the midplane temperature
distributions in the P20, P100 and P1000 models.  The top row has the
radiative equilibrium models, where we neglect the planets' effects on
the pressure scale height (cases P20R, P100R and P1000R).  The bottom
row has the models in both radiative and hydrostatic balance (P20RH,
P100RH and P1000RH).  To highlight the perturbations, we plot the
ratio of perturbed to unperturbed temperature, the latter being as in
Equation~\ref{eq:t_0}.  Figure~\ref{fig:temp_2d} shows the azimuthally
averaged profiles of the temperature and surface density, ratioed with
the unperturbed models.

The biggest disturbance in the temperature comes from the annular gap
in the disk.  Stronger temperature excursions generally go with
stronger density perturbations, that is, with more massive planets.
Among the $h=h^0$ cases, the temperature departs about $\pm$5\% in the
P20R model, and up to $\pm30\%$ in the P1000R model.  The gap is
cooler by about 5\% and 15\% in the P20R and P100R models,
respectively, while temperatures near the gap edges increase by
similar fractions.  In the P1000R model, temperatures both inside the
gap and at its outer edge increase around 25\%, while the gap's inner
edge cools by about 15\%.

The disks that are also in hydrostatic equilibrium have larger
temperature excursions. What's more, the midplane temperature
gradients near the planets change, such that the temperature is
constant at 40~K across the gap in P20RH, while it increases with
radius in P100RH and P1000RH.  Specifically, the temperature rises
from 37~K at 9~AU to 41~K at 13~AU in P100RH, and from 42~K at 7~AU to
56~K at 11~AU in P1000RH.  In the P20RH and P100RH models, the
midplane within the gap is respectively about 5\% and 15\% colder than
the same location in the unperturbed case, while in the P1000RH model,
the gap is about 40\% hotter than the unperturbed disk.  This
difference is due to the fact that the gaps in the P20RH and P100RH
models are optically thick to the starlight, while the gap is
optically thin in the P1000RH model, letting more scattered starlight
and thermal re-emission reach the midplane.  The temperature
variations in the P1000RH model are consistent with the results for
Jupiter presented in \citet{2012ApJ...748...92T}.

The outer edge of the gap opened by the planet is always hotter in the
radiative-hydrostatic equilibrium models P20RH, P100RH and P1000RH
than in the corresponding $h=h^0$ cases, and hotter than the same
location in the unperturbed model.  The higher temperatures mean the
gap outer edges puff up and cast shadows, cooling the disk beyond.  In
the P20RH and P100RH models, the shadows extend to 20 and 30~AU,
respectively, where the decrease in stellar gravity with distance
allows the disk thickness to increase to the point where the upper
layers are once again in starlight.  In the P1000RH model, the shadow
extends beyond 50~AU where our model ends.  In this latter case, the
shadowed regions' weaker stellar illumination means temperatures are
only about 90\% of the unperturbed values, reducing the bolometric
cooling rates roughly by half.

Finally, the temperature variations are slightly asymmetric,
correlating to first approximation with the asymmetries in the surface
density.  For example, in the P1000RH model, the outer edge of the gap
opened by the planet is hotter near the position of the density
enhancement located at about 12~o'clock in figure~\ref{fig:temp_radmc}
lower right panel.  Along this same direction we register steeper
temperature gradients between the hot gap edge and the cold outer
shadowed region.

\subsection{Planet Perturbations' Effects on Panchromatic Disk Continuum Emission}
\label{sec:emission}

We construct synthetic images of all seven models using the
ray-tracing module of RADMC3D, at wavelengths of 1~$\mu$m, 1~mm, and
1~cm.  Whereas the near-infrared is dominated by stellar radiation
scattered by sub-micron grains located in the disk's surface layers,
the emission at millimeter and centimeter wavelengths is mostly
thermal, and originates from larger particles near the midplane.  For
simplicity, we generate synthetic images with the disk face-on.
Furthermore, to better show the planet's effects on the disk's
appearance, in all following figures we ratio each image with that of
the unperturbed model RH.
  
\begin{figure*}[!t]
\centering
\includegraphics[width=1.0\linewidth]{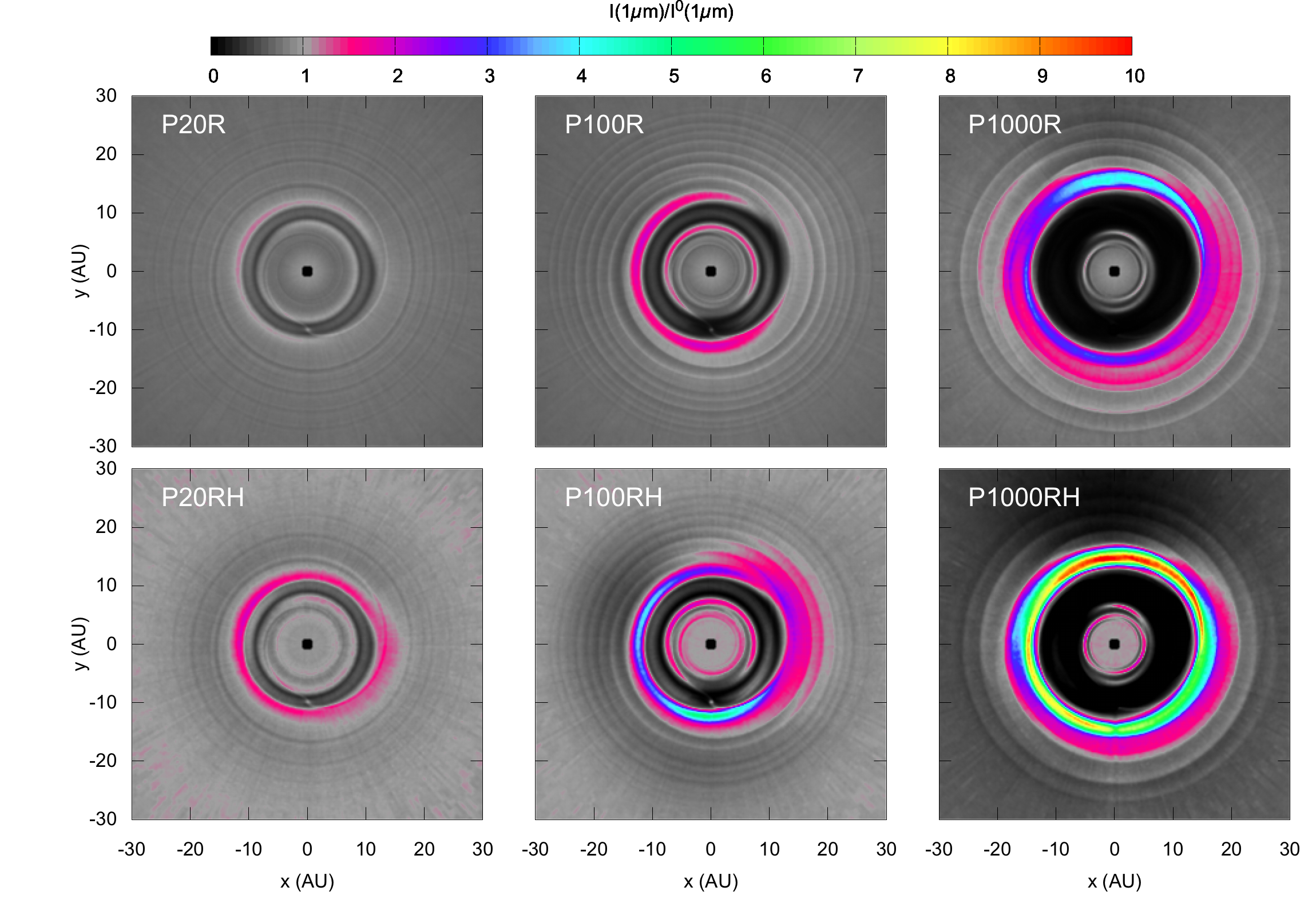}
\caption{Maps of the 1~$\mu$m scattered light for the 20 (left), 100
  (middle), and 1000-M$_\oplus$ cases (right) viewed face-on.  The top
  row shows the models in radiative but not hydrostatic equilibrium,
  P20R, P100R and P1000R.  The bottom row has the corresponding models
  in both radiative and hydrostatic balance, P20RH, P100RH and
  P1000RH. }
\label{fig:1um}
\end{figure*}

\begin{figure*}[!t]
\centering
\includegraphics[angle=0, width=1.0\linewidth]{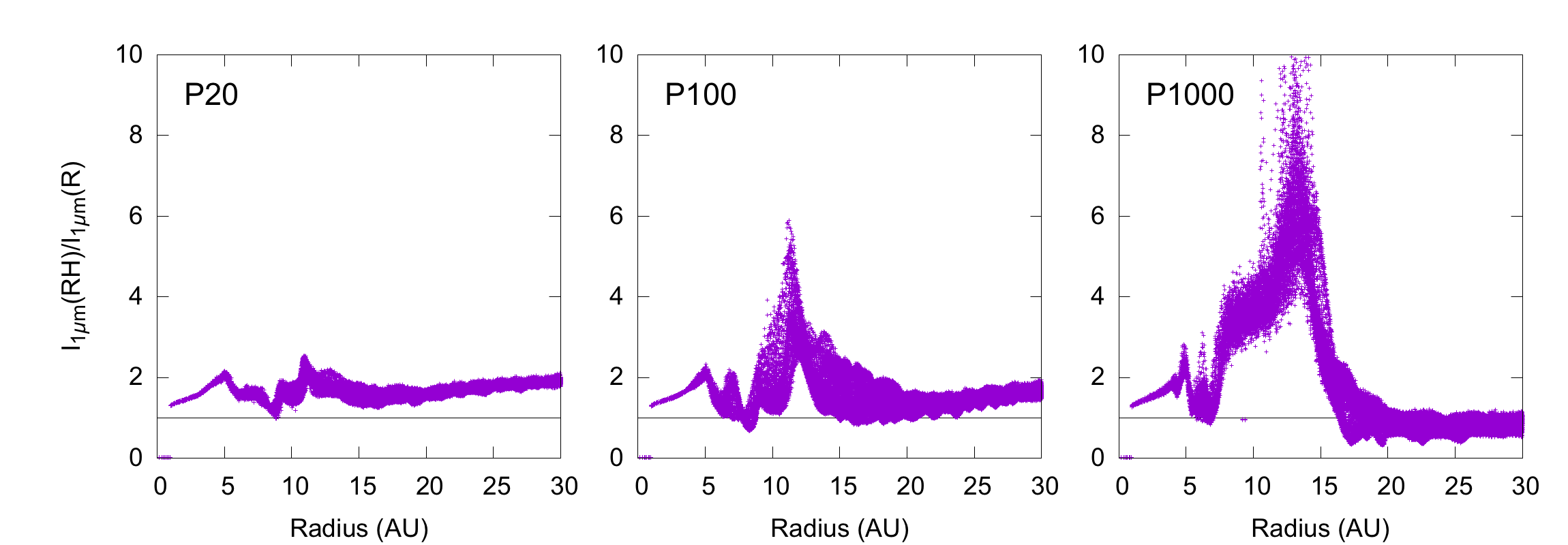}
\caption{Ratio of the scattered light at 1~$\mu$m between the models
  in joint radiative and hydrostatic equilibrium (RH, shown in the
  bottom panels of Figure~\ref{fig:1um}) and those in radiative but
  not hydrostatic equilibrium (R, shown in the top panels of
  Figure~\ref{fig:1um}).  Each point represents one pair of
  corresponding pixels.}
\label{fig:1um_ratio}
\end{figure*}
The 1~$\mu$m synthetic images appear in Figure~\ref{fig:1um}.  Their
strongest features are dark and bright rings tracing, respectively,
the gap opened by the planet and its outer edge.  The hydrostatic
equilibrium models also have a broad dark ring where the gap's
puffed-up outer edge throws its shadow.  The intensity of the bright
ring at the gap's outer edge depends sensitively on the disk's
vertical structure (Figure~\ref{fig:1um_ratio}): its surface
brightness is 1.5-4~times the planet-free case in the three R~models
which are out of hydrostatic balance, and 3-10~times the planet-free
case in the three RH~models, where hydrostatic balance is restored. In
this and other ways, the R- and RH-models appear quite different,
despite having identical surface density maps.

In the hydrostatic models, the region shadowed by the gap's outer edge
scatters starlight with surface brightness just a fraction of the same
locations in the unperturbed disk.  In contrast, the $h=h^0$ R-models
show no such shadowing.  Because of its sensitivity to the scale
height, the scattered light is a poor tracer of the surface density.
Relating scattered light features directly to density variations can
lead to serious errors.

Another notable feature of the scattered light images is that the
planets' outer spiral density waves are brighter in the radiative
equilibrium R-models than in the radiative-hydrostatic equilibrium
RH-models.  For example, in the P1000R model, the spiral arms near
20~AU are 1.5 to 2~times brighter than surrounding regions, a contrast
similar to the amplitude of the spiral density wave.  However in the
corresponding hydrostatic P1000RH model, the spiral arms disappear in
the dark ring beyond the gap's outer edge.  We will come back to this
point in Section~\ref{sec:disc} in comparing models to observations.

\begin{figure*}[!t]
\centering
\includegraphics[ width=1.0\linewidth]{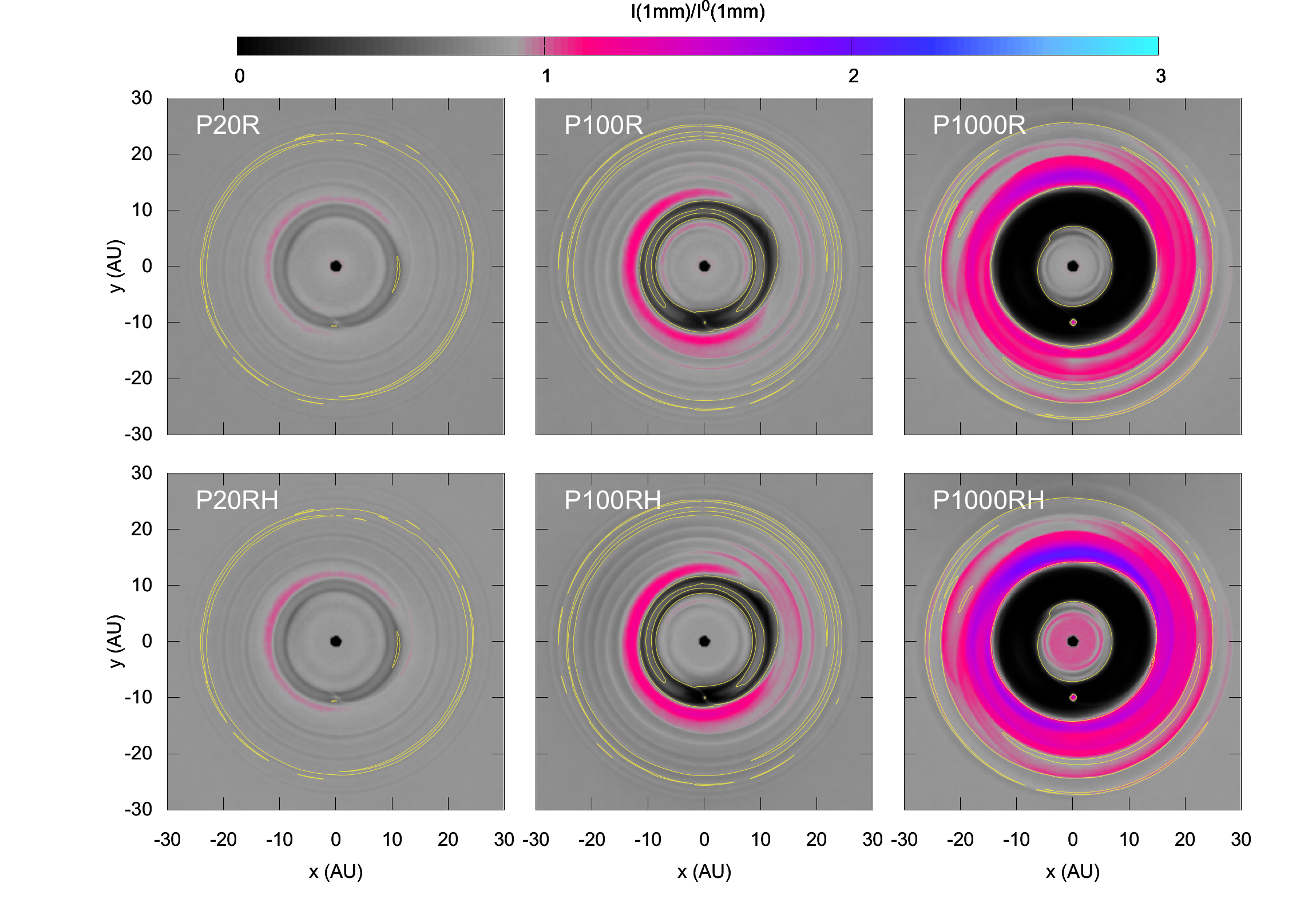}
\caption{ Maps of the 1~mm continuum emission from cases with planets
  of 20 (left), 100 (middle), and 1000~M$_\oplus$ (right) viewed
  face-on.  The top row shows the radiative equilibrium models P20R,
  P100R and P1000R.  The bottom row has the radiative-hydrostatic
  models P20RH, P100RH and P1000RH.  Yellow contours mark
  monochromatic optical depth unity.  }
\label{fig:1mm}
\end{figure*}

\begin{figure*}[!t]
\centering
\includegraphics[width=0.9\linewidth]{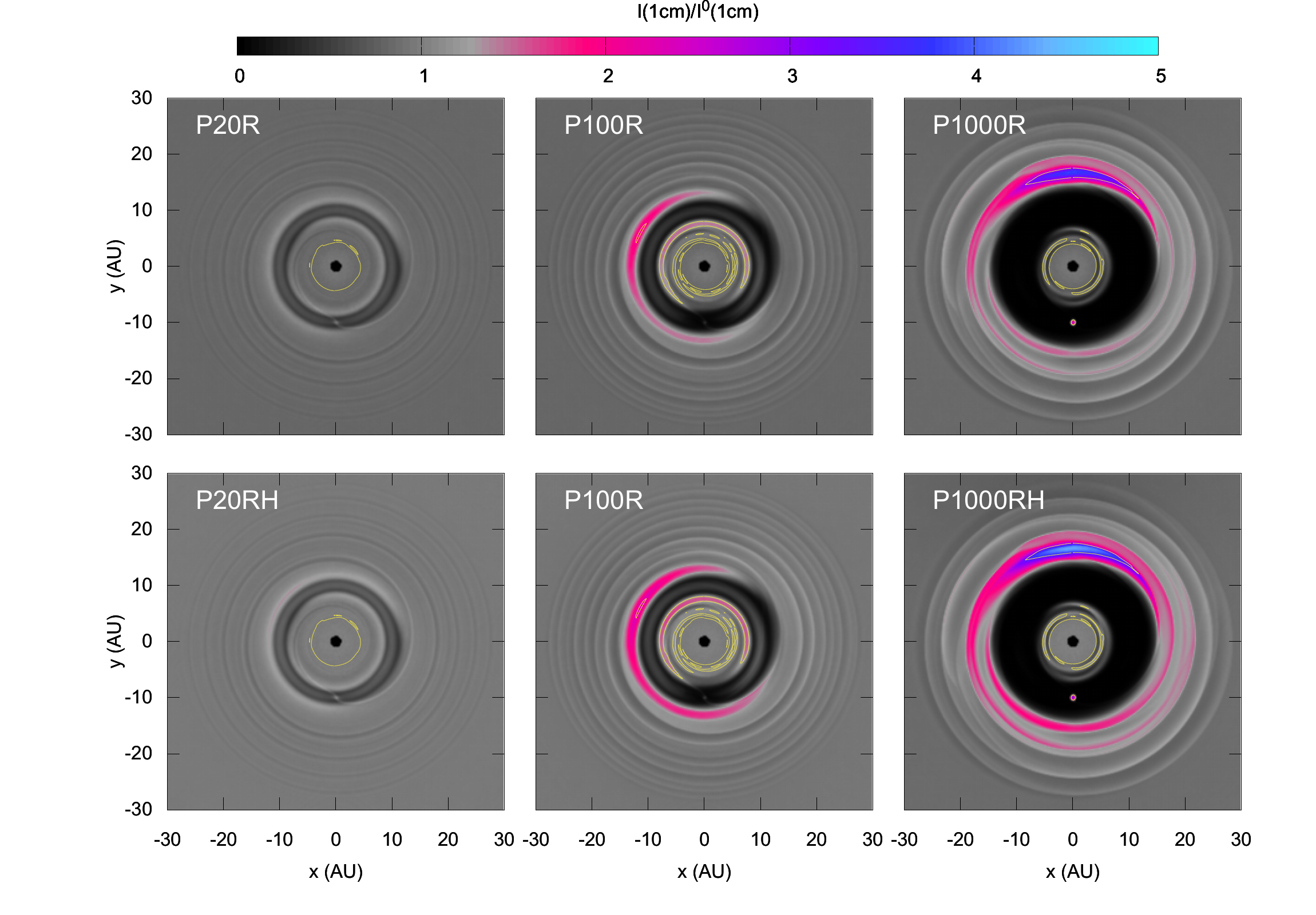}
\caption{As Figure~\ref{fig:1mm} but for the wavelength of 1~cm.}
\label{fig:1cm}
\end{figure*}
 
Figures~\ref{fig:1mm} and \ref{fig:1cm} show the ratio between the
dust emission of the perturbed and unperturbed disk models at
wavelengths 1~mm and 1~cm, respectively.  As in the scattered light,
the gaps opened by the planets appear as dark annuli, and the gaps'
outer rims appear as bright annuli.  Azimuthal asymmetries due to
vortices orbiting near the gaps' outer edges are also visible.

For our assumed dust opacity and initial surface density, the 1~mm
emission arising within 20~AU of the central star is mostly optically
thick, and so probes spatial variations of the temperature rather than
the surface density.  For example, in the P20 models, the dark ring at
the partially depleted, but still optically thick, gap comes from the
low temperature in the gap discussed in Section~\ref{sec:temp} and is
not due to the lower surface density.  The large optical depth also
reduces the visibility of the azimuthal asymmetries in the P100 and
P1000 models. In both the P1000R and RH models, the intensity ratio at
the center of the vortex on the gap's outer edge is about 2.5, which
is only about half the surface density ratio (Figure~\ref{fig:fargo}).

Since the 1~mm emission probes temperatures, it indirectly informs us
about the vertical structure of the disk.  In particular, the
radiative-hydrostatic equilibrium RH-models have brighter gap outer
edges, and fainter shadowed regions, than the
radiative-equilibrium-only R-models.  The starlight illumination is
thus important for interpreting millimeter-wave observations of the
optically-thick parts of protostellar disks.

The dust thermal emission at the wavelength of 1~cm from regions more
than 5~AU from the star is mostly optically thin.  It more closely
probes the surface density perturbations caused by the planets.  For
example, at the gap's outer edge the intensity varies with azimuth by
factors of about 1.3 in the P20RH model, and about 5 in the P1000RH
model, similar to the azimuthal ranges in the surface density shown in
Figure~\ref{fig:fargo}.  As with the 1~mm emission, the radiative
equilibrium and radiative-hydrostatic equilibrium models look
different at 1~cm wavelength, following the corresponding variations
in the dust temperature discussed in the previous section.

\section{Consequences for Gas and Planet Dynamics}

The main goal of our investigation is to understand young planets'
effects on the multi-wavelength dust continuum emission from
protostellar disks, especially at the locations outside 5~AU where
starlight dominates the heating.  In pursuing this, we have found that
the planet-disk interaction produces more than the long-predicted
surface density perturbations: it also significantly perturbs
temperatures across the disk.  In this section, we elaborate on the
temperature perturbations' consequences for the dynamics of gas, dust
and planets.

\subsection{Heating and Cooling}

\begin{figure*}[!t]
\centering
\includegraphics[angle=0, width=1\linewidth]{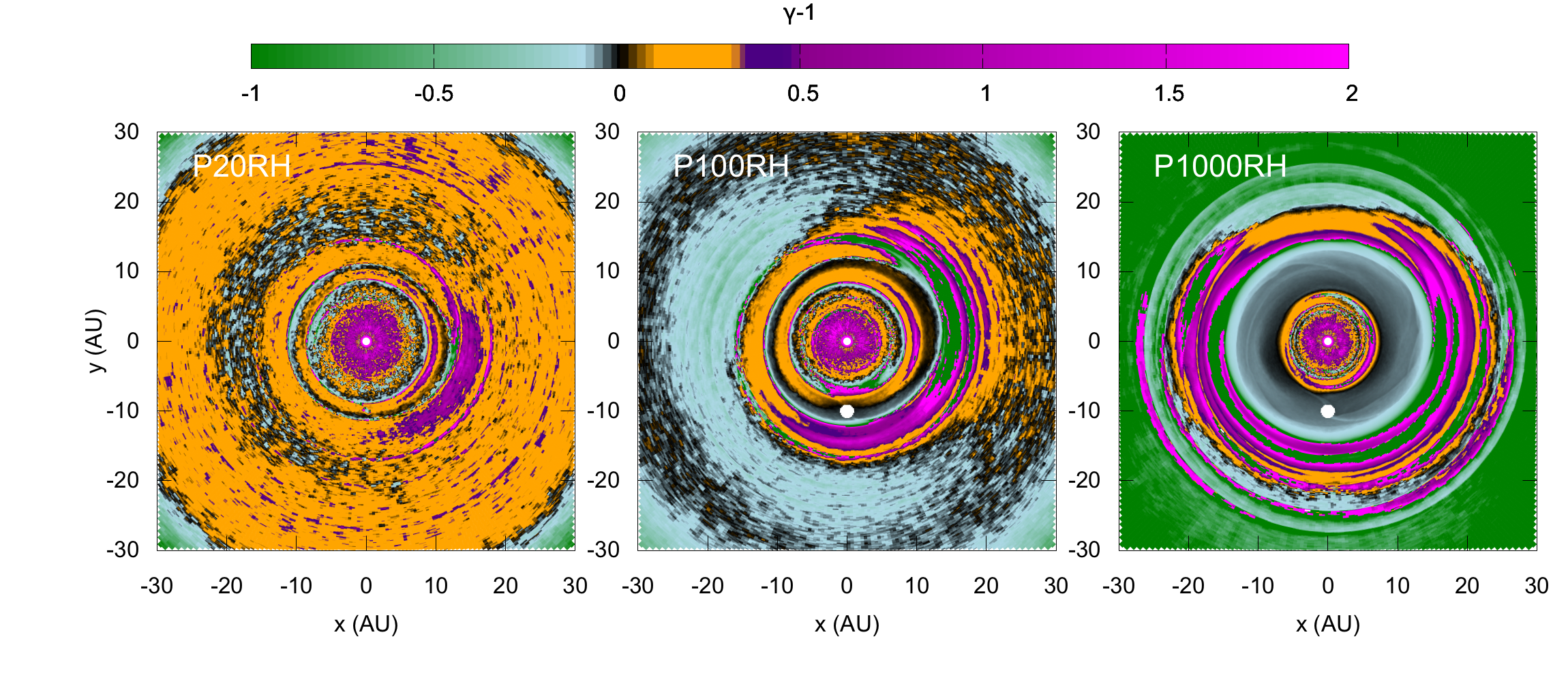}
\caption{Maps of the power-law index in $T\propto\rho^{\gamma-1}$
  describing the pointwise change in midplane temperature and density
  between the disk's unperturbed state and its radiative-hydrostatic
  equilibrium with an embedded planet.  The three panels show models
  P20RH, P100RH and P1000RH (left to right).}
\label{fig:gamma}
\end{figure*}

We showed in Section~\ref{sec:temp} that perturbing planets of more
than about 20~$M_\oplus$ yield variations in the disk temperature with
respect to a planet-free disk.  This means that evolving the disk with
an isothermal equation of state produces errors, and that reliable
hydrodynamical modeling requires treating the heating and cooling.
The need for something beyond a simple power-law relationship between
temperature and density is illustrated in Figure~\ref{fig:gamma}, a
map of the index $\gamma-1$ in the relationship
$T(r,\phi)\propto\rho^{\gamma-1}(r,\phi)$ applied to the disk's
evolution from the unperturbed state to the planet perturbation in
radiative-hydrostatic equilibrium.  The index is calculated by
\begin{equation}
  \gamma-1 = \log(T_m^{RH}/T_m^0) / \log(\rho_m^{RH}/\rho_m^0),
  \label{eq:gamma}
\end{equation}
where $T_m^0$ is the midplane temperature of the unperturbed disk and
$T_m^{RH}$ is the midplane temperature in the RH-model.  Isothermal
evolution corresponds to $\gamma-1 = 0$, and adiabatic evolution of
molecular hydrogen corresponds to $\gamma-1 = 0.4$.  Clearly the
evolution from the unperturbed to planet-perturbed models is not well
described by a single value of $\gamma$.  Note that $\gamma-1$ is near
zero within the gap opened by the planet not because the temperature
is time-constant, but because the change in density is so much greater
than the change in temperature.  Generally, $\gamma-1$ ranges from
positive to negative values as we move from patches of the disk
directly lit by the star to those lying in the shadows.

\begin{figure*}[!t]
\centering
\includegraphics[angle=0, width=1\linewidth]{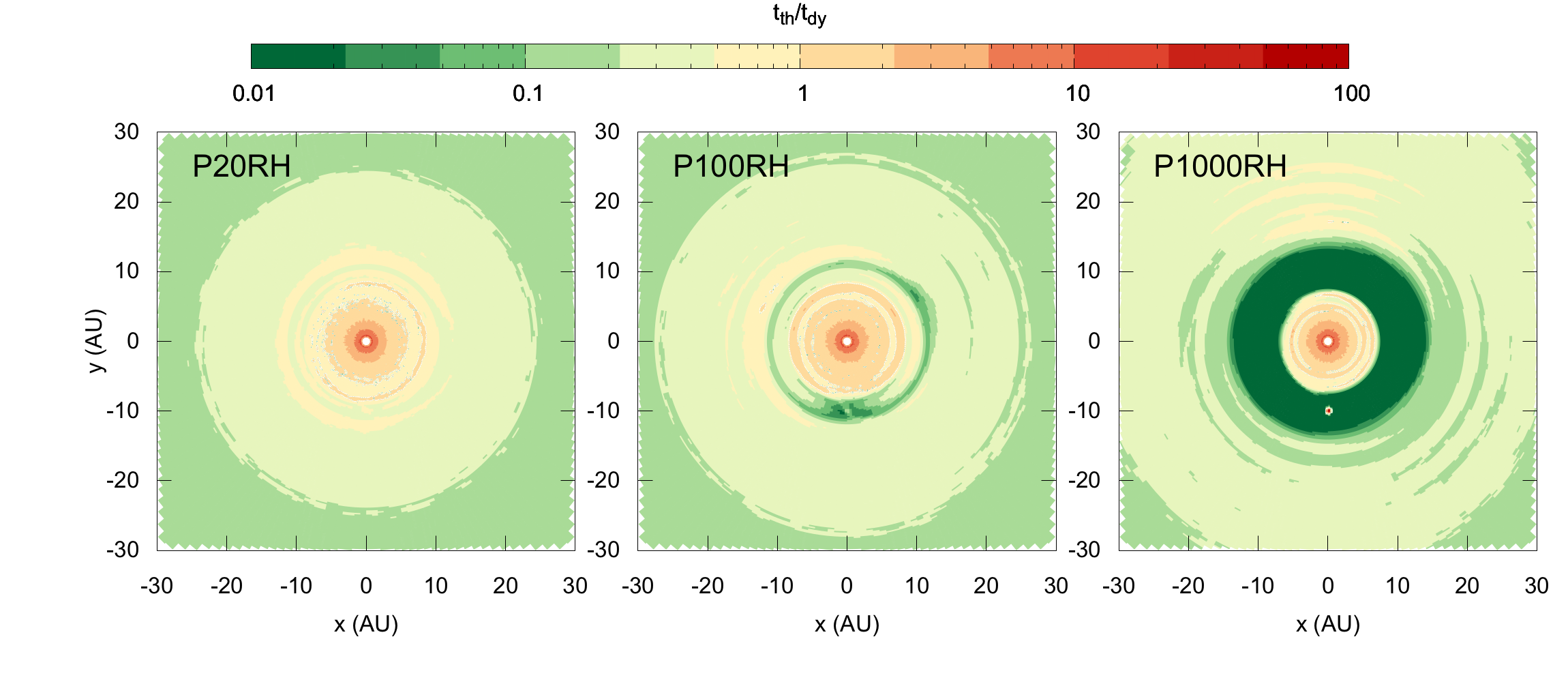}
\caption{Ratio between the thermal and dynamical time scales.  The
  three radiative-hydrostatic equilibrium models P20RH, P100RH and
  P1000RH appear from left to right.  }
\label{fig:time_scale_ratio}
\end{figure*}

Another consequence of the changes in surface density and temperature
across the disk is new values for the thermal time scale discussed in
section~\ref{sec:disk_temp_mod}.  In figure~\ref{fig:time_scale_ratio}
we show the ratio of the thermal to the dynamical time scale in the
three radiative-hydrostatic equilibrium cases.  The 20-M$_\oplus$
model is broadly similar to the unperturbed profile in
figure~\ref{fig:timescale}.  The 100-M$_\oplus$ model cools faster in
the gap owing to the lower surface density, and cools slower in the
outer shadowed region owing to the low temperatures.  The
1000-M$_\oplus$ model is a more extreme version of the 100-M$_\oplus$
case, with shadows so deep and cold that temperatures change only over
time scales comparable to or longer than the local orbital period.
Apparently gap-opening planets can drastically change the surrounding
disk's thermal timescale, and thus its response to changing
illumination.

\subsection{Orbital Migration}

Temperature perturbations caused by the planet-disk interaction might
also affect the planet's long-term orbital migration, which is
governed by the gravitational forces exchanging angular momentum
between planet and disk.  If $M_p\ll M_{th}$, the rate and direction
of migration are sensitive to the disk's temperature and density
structure.  In a power-law model disk defined by $T\propto r^{-\beta}$
and $\Sigma\propto r^{-s}$, with $\beta$ between 0.1 and 0.5, and $s$
between 1 and 1.5, an Earth-mass planet migrates inward at a rate as
high as $10^{-5}$~AU/yr \citep{1997Icar..126..261W,
  2006RPPh...69..119P}.  However, sufficiently steep radial
temperature gradients can slow and even reverse the motion.  In
particular, outward migration is expected within a few~AU of the star,
where accretion is the main source of heating, while inward migration
should occur at larger radii \citep{2006A&A...459L..17P,
  2010ApJ...715L..68L, 2011MNRAS.410..293P, 2011A&A...536A..77B}.  Our
results suggest that planets less than $M_{th}$ perturb the
temperatures too little to affect the migration rate.  However,
low-mass planets could find themselves migrating along the temperature
gradients arising from a more massive companion's interaction with
their shared host disk.

Temperature perturbations' effects on planets with $M_p\sim M_{th}$
have received too little attention.  \citet{2005ApJ...619.1123J} found
that the density and temperature perturbations near Neptune-mass
planets can substantially slow their inward migration.  Similar
results were found by \citet{2003ApJ...599..548D}, who moreover argued
that the lack of an appropriate energy equation is the main limitation
in modeling such planets' dynamics.  Our results show that planets
with close to the thermal mass (as in the P20 models) produce
disturbances of about $\pm$10\% in the gas temperature
(Figure~\ref{fig:temp_2d}).  More importantly, in the P20RH model, the
temperature gradient within the planet's partially depleted gap is
close to zero.  Quantifying the effects such perturbations have on the
migration requires numerically integrating the torque exerted by the
disk on the planet, which is beyond the scope of this work.  Instead,
we seek a qualitative understanding of the effects on low-mass
planets' migration by assuming the total torque $\Gamma_{tot}$ is the
sum of the Lindblad torque $\Gamma_L$ and the horseshoe corotation
torque $\Gamma_{hs}$ expressed as \citep{2011MNRAS.410..293P,
  2013A&A...549A.124B, 2014prpl.conf..667B}
\begin{eqnarray}
  \Gamma_{tot}  & =  & \Gamma_L + \Gamma_{hs}, \label{eq:torque1}\\
  \gamma\frac{\Gamma_L}{\Gamma_0} & = & -2.5 - 1.7\beta + 0.1s, \\
  \gamma\frac{\Gamma_{hs}}{\Gamma_0} & = & \frac{7.9}{\gamma}(\beta-(\gamma-1.0))s, \label{eq:torque3}\\
  \Gamma_0 & = & \left( \frac{q}{h} \right)^2 \Sigma_p r_p^6 \Omega_p^2, \label{eq:torque4}
\end{eqnarray}
where $\gamma$ is the effective ratio of specific heats for the
disturbances the planet launches, taking into account radiative
diffusion.  If the disturbances are so optically thick they trap their
thermal radiation, then they propagate through the disk in the
adiabatic limit, with $\gamma=1.4$ for molecular hydrogen gas.  If the
photons quickly smooth out temperature variations, then the
disturbances propagate isothermally, and $\gamma=1$.  In the models we
consider, the thermal timescale at the planet is about half the
orbital period, so the thermal diffusivity is $2 h^2\Omega/(2\pi)$.
The effective equation of state considering disturbances of all
wavenumbers is then around $\gamma=1.2$, judging from the numerical
results shown in figure~8 of \citet{2011MNRAS.410..293P}.  This
implies that the total torque in the unperturbed model ($s=1$,
$\beta=0.45$) is negative with $\Gamma_{tot}/\Gamma_0=-1.3$,
indicating inward migration as expected.  In model P20RH, the surface
density near the planet's orbit also varies as $r^{-1}$, while the
temperature is roughly constant between 9~AU and 11~AU, an annulus
that spans both the horseshoe region and the Lindblad resonances'
pileups inside and outside the planet's orbit, and thus most strongly
affects the planet.  The uniform temperature yields a more negative
torque $\Gamma_{tot}/\Gamma_0=-3.1$, indicating faster inward
migration.

Finally, planets much more massive than $M_{th}$ clear gaps, and so
are thought to migrate following the global inflow of the
circumstellar gas.  The massive planets through their temperature
perturbations might not affect their own migration rate, but they can
alter the migration of nearby low-mass planets.  For example, the
P1000RH model in figure~\ref{fig:temp_radmc} at 15-20~AU, in the
shadow of the gap's outer edge, has midplane temperature scaling as
$T\propto r^{-1.6}$.  Across the same annulus, the surface density
varies as $\Sigma\propto r^{-3.9}$.
Equations~\ref{eq:torque1}-\ref{eq:torque3} above indicate that a
low-mass planet ($M_p\ll M_{th}$) orbiting here feels a positive
torque ($\Gamma_{tot}/\Gamma_0\sim 26$) so migrates rapidly outward.
Positive torques also occur in the shadowed regions of the P20RH and
P100RH models.  Further out in each shadow, the temperature and
surface density gradients flatten, and the torque turns negative.
Migrating low-mass planets thus likely converge within the shadows
cast by the gaps opened by the massive planets.

\section{Consequences For Interpreting Features in Protostellar Disks}
\label{sec:disc}

\subsection{Taxonomy}

This section is devoted to the models' implications for interpreting
recent observations.  Many protostellar disks feature annuli,
crescents, and circular and spiral arcs in the dust and gas emission,
as revealed through the improved sub-arcsecond mapping capabilities of
infrared cameras such as HiCIAO/Subaru, Sphere/VLT, GPI/Gemini, and
millimeter and centimeter arrays including CARMA, SMA, ALMA and the
VLA.  A list of such disks is in Table~\ref{tab:disks}.  Whether these
structures, and their diversity, result from the interaction between
the circumstellar material and forming planets, or are caused by other
processes, is debated.

\begin{figure*}[!t]
\centering
\includegraphics[width=0.7\linewidth]{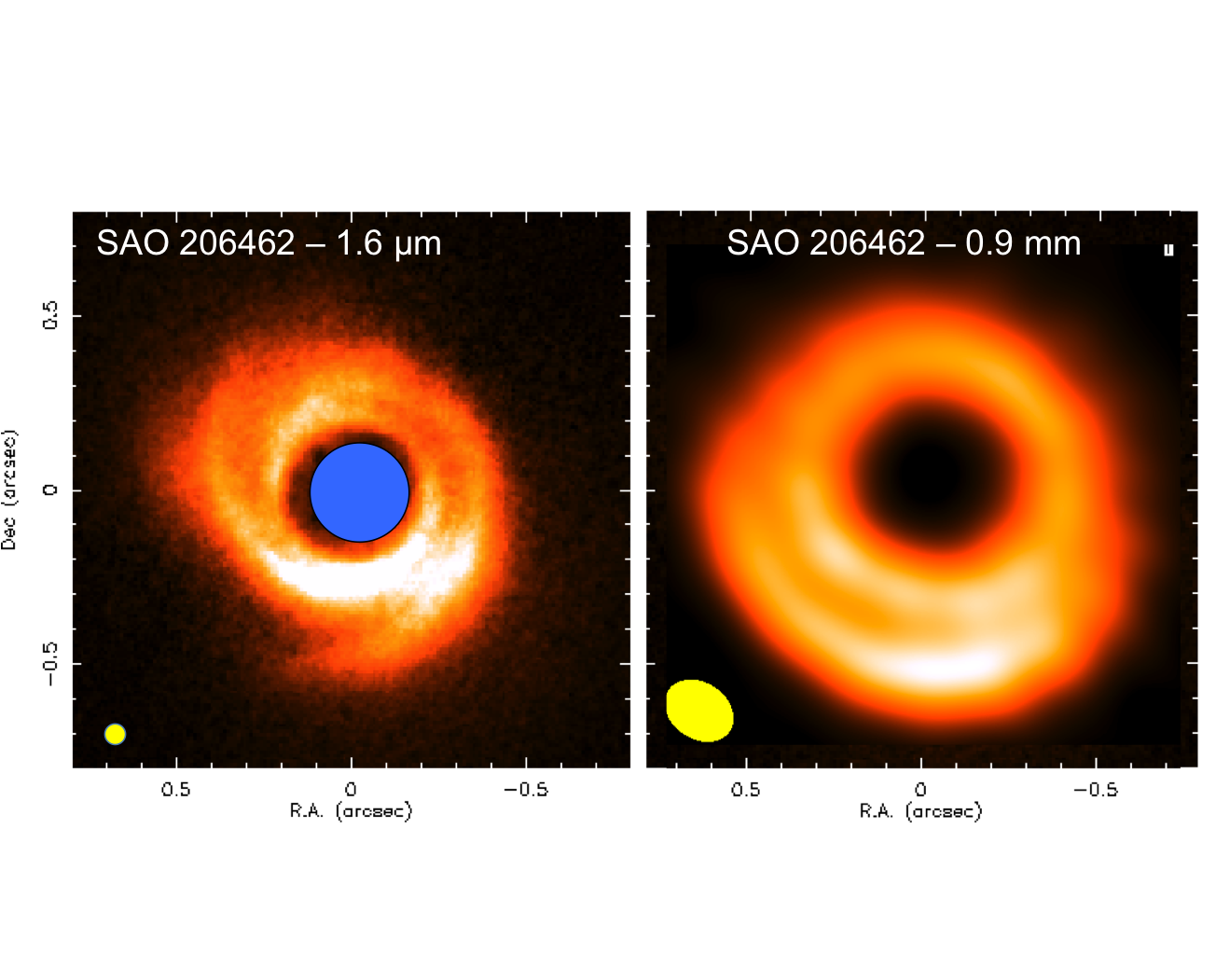}
\includegraphics[width=0.7\linewidth]{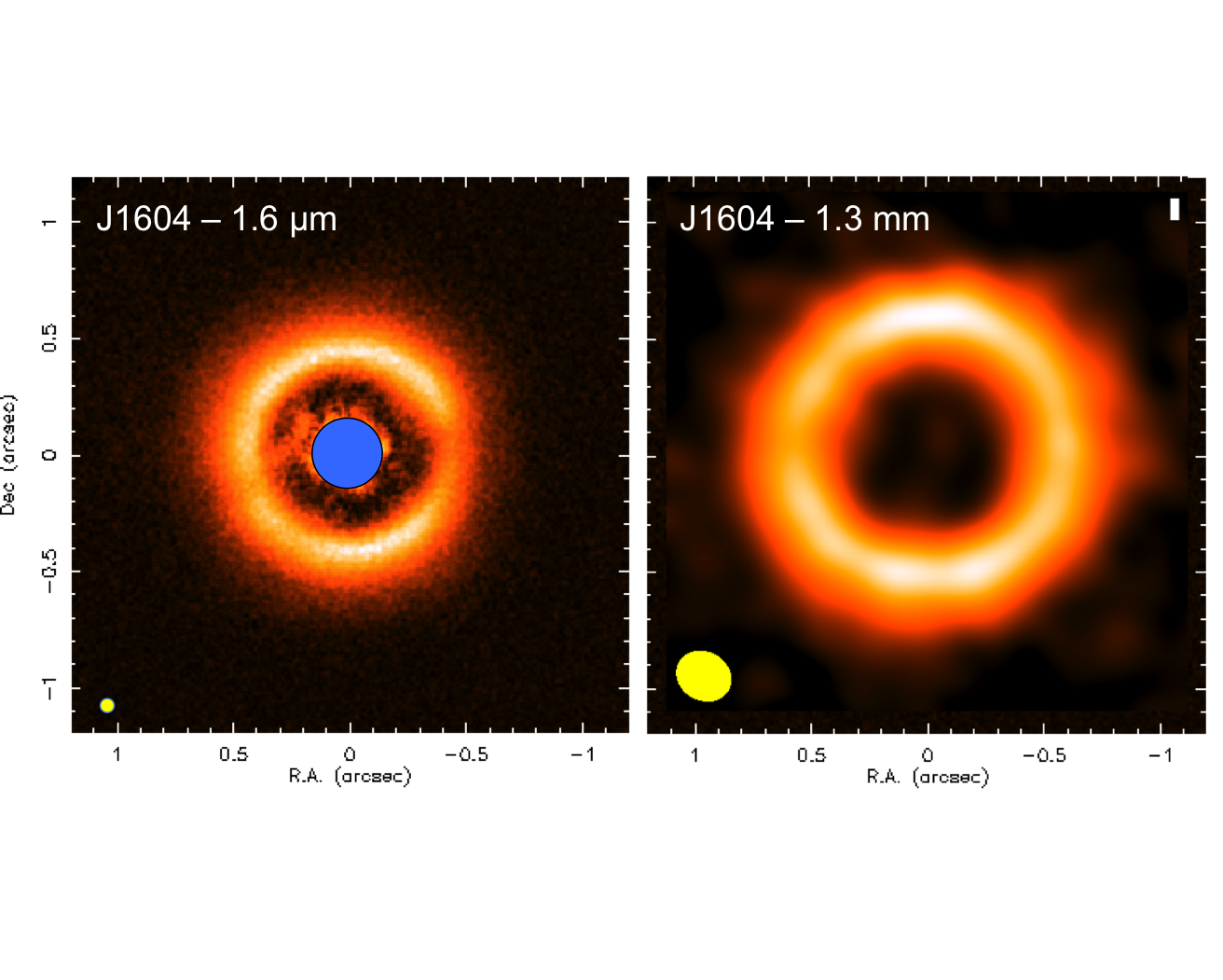}
\caption{Maps of the SAO~206462 (top) and J1604 (bottom) disks
  recorded in the continuum emission at 1.6~$\mu$m (left) and
  $\sim$1~mm (right).  These maps illustrate the diversity of
  structures observed in protostellar disks, which include spiral and
  circular arcs, as well as crescents.  The observations are taken
  from \citet{Muto2012}, \citet{Mayama12}, \citet{2017ApJ...836..201D}
  and \citet{2016ApJ...832..178V}.}
\label{fig:disks}
\end{figure*}

The SAO~206462 and J160421.7-213028 systems are good examples of the
diverse morphologies observed in protostellar disks
(Figure~\ref{fig:disks}).  At sub-millimeter wavelengths, both disks
show partially depleted dust cavities tens of AU in diameter
\citep{Perez2014, Zhang2014, 2016ApJ...832..178V,
  2017ApJ...836..201D}.  However, while the dust continuum in the
SAO~206462 disk traces a ring and a crescent, the J1604 disk appears
almost axisymmetric.  At near-infrared wavelengths, the SAO~206462
disk exhibits an $m=2$ grand design spiral, while the J1604 disk
appears as a circularly symmetric annulus.

The crescents observed in the microwave regime come with a large
spread of amplitudes, defined as the difference between the maximum
and minimum flux densities measured at the same orbital radius divided
by the minimum intensity, $a = (F_{max}(r)-F_{min}(r))/F_{min}(r)$.
IRS~48 has the most prominent crescent observed so far, with an
amplitude greater than $100$; MWC~758, SAO~206462, LkHa~330 and SR~21
have much more moderate amplitudes between~1.5 and~3.  Recent data
show that crescents at sub-millimeter wavelengths are accompanied in
most cases by near-infrared spiral features.  This is the case for
AB~Aur, MWC~758, SAO~206462 and HD~142527.

The observations collected so far reveal a second trend: the earlier
the spectral type of the star, the more complex the morphology of its
disk.  For example, the disks around HD~100546 and AB~Aur, which have
spectral types of B9.5 and A0, respectively, show multiple spiral
features in the near-IR.  The AB~Aur disk has a complex morphology
also at millimeter wavelengths, showing a crescent and spiral features
in the molecular line emission \citep{Tang2012}.  The disks around
MWC~758, SAO~206462, and HD~142527, with spectral types between A5 and
F7, have $m=1$ or $m=2$ spiral structures. The disks around HD~97048
(A0), HD163296 (A1), HD~169142 (A5) are exceptions in that they do not
show spiral features. In contrast, the disks around late G, K and M
stars are mostly characterized by azimuthally symmetric rings at both
infrared and millimeter wavelengths.


Finally, we note that in classifying perturbed disks, we explicitly
avoid making use of the morphology of the molecular line emission
observed with ALMA.  The main reason is that spatially-resolved maps
of the molecular line emission are available only for a few of the
disks listed in Table~\ref{tab:disks}.  The molecular line emission
however provides valuable information on the nature of the structures
observed in disks.  For example, ALMA observations of HD~142527,
SAO~206462, and MWC~758 show that the $^{13}$CO and C$^{18}$O are
distributed much more nearly axisymmetrically than the dust.  This
suggests the crescent observed in the millimeter-wave continuum could
result from the decoupling of large dust grains from the gas.

\begin{table}
 \caption{Protostellar disks around stars not known to be double or
   multiple, and which show small-scale morphological features.  The
   list includes only objects with high-quality imaging at both
   near-infrared (NIR) and millimeter wavelengths (mm).  The source
   list is ordered by stellar spectral type.  In Columns 4 and 5, we
   report the main morphological feature characterizing NIR and mm
   observations, respectively.  The parameter $m$ indicates the number
   of structures.  The parameter $a$ indicates the azimuthal contrast
   in the dust continuum emission at the crescent(s)' orbital radius.}
 \label{tab:disks}
 \begin{tabular}{llllll}
 (1) & (2) & (3) & (4) & (5) & (6) \\
Name  		  	& Sp.T. & M$_\star$  & NIR 		& mm 			& ref. \\
\hline
\hline 
HD~100546		& B9.5	& 2.4	& Spiral, $m \geq 3$	& Ring, $m \geq 1$ 				& \tablenotemark{1} \\
HD~97048                & A0  & 2.5 & Ring, $m \geq 4$                             & Ring , $m \geq 2$               & \tablenotemark{2}\\
AB~Aur			& A0	& 2.0	 & Spiral, $m \geq 5$  	& Crescent, $m\geq1$, $a \sim 4$	& \tablenotemark{3} \\
IRS~48			& A0	& 2.0	 & Spiral, $m \geq 1$		& Crescent, $m\geq1$, $a > 100$	& \tablenotemark{4} \\
HD~163296              & A1 & 2.3 & Ring, $m \geq 1$		& Ring, $m \geq 3$	     	& \tablenotemark{5} \\
MWC~758	   	& A5	& 2.0	& Spiral, $m \geq 2$		& Crescent, $m\geq2$, $a \sim 1.5, 3$	& \tablenotemark{6} \\
HD~169142		& A5	& 2.0	& Ring, $m \geq 2$		& Ring, $m\geq2$			& \tablenotemark{7} \\
SAO~206462	    	& F4	& 1.7	& Spiral, $m \geq 2$ 	& Ring, $m\geq1$; Crescent, $m\geq1, a \sim 3$	& \tablenotemark{8} \\
LkH$\alpha$~330	& G3	& 2.5	& Spiral, $m \geq 1$		& Crescent, $m\geq1, a \sim1.5$	& \tablenotemark{9} \\
SR~21			& G3	& 2.5	& No structures   		& Crescent, $a \sim 3$ 	& \tablenotemark{10} \\
J160421.7-213028	& K2	& 1.0	& Ring, $m\geq1$				& Ring, $m\geq1$ 				& \tablenotemark{11} \\
LkCa~15			& K3	& 1.0	& Ring, $m\geq2$		& Ring, $m \geq 1$				& \tablenotemark{12} \\
RX~J1615-3255	& K5	& 1.1	& Ring, $m\geq3$		& Ring, $m \geq 1$				& \tablenotemark{13} \\
PDS~70			& K5	& 0.8	& Ring, $m\geq1$		& Crescent, $ a \sim 1.5$	& \tablenotemark{14} \\
TW~Hya			& K6 & 0.8     & Ring, $m \geq 5$			& Ring, $m\geq 6$ 		& \tablenotemark{15} \\
\hline
\end{tabular}
\tablenotetext{1}{\citet{2001AJ....122.3396G, 2007ApJ...665..512A, 2013ApJ...766L...1Q, 2014ApJ...791L...6W, 2016A&A...588A...8G}}
\tablenotetext{2}{\citet{2016A&A...595A.112G, 2017A&A...597A..32V}}
\tablenotetext{3}{\citet{2011ApJ...729L..17H, Tang2012}} 
\tablenotetext{4}{\citet{2015A&A...579A.106V, 2015ApJ...798..132F}}
\tablenotetext{5}{\citet{2014A&A...568A..40G, Isella2016, 2016ApJ...818L..16Z}}
\tablenotetext{6}{\citet{Benisty2015, Isella10b, 2017arXiv171208845B}}
\tablenotetext{7}{\citet{2013ApJ...766L...2Q, 2015PASJ...67...83M, 2018MNRAS.473.1774L,2017ApJ...850...52P,2018MNRAS.474.5105B,2017ApJ...838...97M}}
\tablenotetext{8}{\citet{Muto2012, Perez2014,2017ApJ...849..143S, 2016ApJ...832..178V}}
\tablenotetext{9}{\citet{2016arXiv160704708A, Isella13}}
\tablenotetext{10}{\citet{2013ApJ...767...10F, Perez2014, 2015A&A...579A.106V}}
\tablenotetext{11}{\citet{2012ApJ...760L..26M, Zhang2014, 2015A&A...584L...4P, 2017ApJ...836..201D}}
\tablenotetext{12}{\citet{Andrews11b, Isella12, 2014ApJ...788..129I,  2014A&A...566A..51T, 2016ApJ...828L..17T}}
\tablenotetext{13}{\citet{ 2016A&A...595A.114D,2015A&A...579A.106V}}
\tablenotetext{14}{\citet{2012ApJ...758L..19H, 2013ApJ...775L..33H, 2015ApJ...799...43H}}
\tablenotetext{15}{\citet{2015ApJ...815L..26R, 2016ApJ...820L..40A,2017ApJ...837..132V}}

\end{table}

\subsection{Comparing Models With Observations}

\noindent {\it Annuli:} The models discussed in
Section~\ref{sec:emission} show that the disk-planet interaction can
naturally lead to ring-like structures in both the infrared scattered
light and millimeter-wave emission, similar to those observed toward
dynamically-cold disks.  For example, the annular emission observed
toward J160421.7-213028 is very similar to the synthetic maps for the
P1000 model shown in the bottom-right panels of Figure~\ref{fig:1um}
and \ref{fig:1mm}.  In both the model and the observations, the
scattered light emission is confined to a narrow ring, whereas dust is
present on a much wider area \citep{Zhang2014, 2017ApJ...836..201D}.
In the case of HD~169142, polarimetric images obtained at about
0.7~$\mu$m show a narrow bright ring with a radius of 23~AU,
surrounded by a wide dark annulus extending to about 65~AU and a
second wide bright ring extending outward of 65~AU
\citep{2018MNRAS.474.5105B}.  These structures are also very similar
to those observed in our models, where the dark ring might trace the
shadow cast by the outer rim of the gap created by a planet. In our
model, rings seen in scattered light are narrow because scattered
light arises from the puffed-up outer edge of the gap opened by the
planet.  If, instead, the disk vertical structure were unaffected by
the planet, scattered light would come from a wider annulus as shown
in the top-right panel of Figure~\ref{fig:1um}.  Clearly the vertical
structure is a key factor in interpreting scattered light maps of
protostellar disks.

The effect of the vertical structure on the scattered light emission
can be understood by analogy with a mountain illuminated by the Sun at
sunset, as seen by an observer flying over it.  Only the part of the
mountain facing west will be illuminated, while its east side and the
surrounding valleys will be in the dark.  The shape of the illuminated
side will depend on the shadows cast by the terrain westward of the
mountain, but, in general, the higher the mountain, the larger the
fraction lit by the setting Sun.  The outer edge of the gap opened by
the planet is similar to such a mountain.  Indeed, this region is
hotter, and therefore more puffed-up, than the surrounding area
because it is directly illuminated by the star
(Section~\ref{sec:temp}).
 
\noindent {\it Crescents:} In an inviscid or low viscosity disk, the
disk-planet interaction produces vortices near the outer rim of the
gap opened by the planet.  This could naturally lead to the crescents
observed in the millimeter- and centimeter-wave dust continuum
emission, if the vortices concentrate sub-millimeter and
millimeter-sized particles through gas drag forces.  The strongest
concentrations toward gas pressure maxima are expected to occur for
particles whose aerodynamic stopping time is comparable to the orbital
period, so that the Stokes number $St\sim 1$
(Section~\ref{sec:dust-gas}).

The crescents observed in perturbed disks are characterized by
azimuthal variations in the flux density with amplitudes between 1.5
and more than 100 (Table~\ref{tab:disks}).  Even if the dust emission
is optically thin, and therefore traces the dust column density,
measuring the amplitude of the crescent is by itself insufficient to
determine the degree to which the dust has been concentrated relative
to the gas.  We must also measure the azimuthal variation in the gas
density.  Nevertheless, hydrodynamic models indicate that a single
planet generates azimuthal perturbations on the gas surface density of
up to a factor of a few (Figure~\ref{fig:fargo}).  Similar values are
obtained with multiple planets \citep[e.g.][]{Isella13}.  Therefore,
the large-amplitude crescents observed in HD~142527 (amplitude~30) and
IRS~48 (amplitude~$>100$) require dust and gas be dynamically
decoupled.

The degree of separation of dust and gas can be measured by comparing
maps of the optically-thin dust and trace molecular species' emission.
The latter is the best probe of the total gas mass, since the main
component, molecular hydrogen, is not itself observable.  In the case
of the HD~142527 circumbinary disk, \cite{2017ApJ...840...60B} find
that the large crescent observed in the dust emission corresponds to
an azimuthal variation of a factor of~54 in the density of millimeter
grains, but only a factor of about~4 in the density of $^{13}$CO and
C$^{18}$O molecules.  While the conversion from CO to H$_2$ density is
hampered by uncertainties in the molecular abundance \citep[see,
  e.g.,][]{2000ApJ...544..903W, 2002A&A...386..622A,
  2015ApJ...813..128Q,2017A&A...599A.113M}, this result suggests that
vortices concentrate dust efficiently, and that even the more
prominent dust crescents might be compatible with azimuthal variations
in the gas density such those predicted by planet-disk interaction
models.
     
The disks surrounding SAO~206462
\citep[fig.~\ref{fig:disks};][]{2016ApJ...832..178V} and MWC~758
\citep{2017arXiv171208845B} show complex combinations of rings and
crescents.  In the first case, an azimuthally extended dust crescent
is observed outside a ring, which itself shows azimuthal variations in
the millimeter-wave dust emission.  In the second case, the 0.87~mm
dust continuum emission shows two dust crescents centered about 47 and
82~AU from the central star, and characterized by azimuthal variations
in the dust emission by factors of 1.5 and~2, respectively.
Qualitatively, these features resemble the structure of the P100RH and
P1000RH models, where dust crescents form at both edges of the gap
formed by a planet.  If this model applies, the perturbing object
might be located at a radius of about 65~AU in the case of MWC~758,
and about 55~AU in the case of SAO~206462.
  

\noindent {\it Spirals:} Interpreting protostellar disks with grand
design spirals in the framework of planet-disk interaction is
problematic.  Our models suggest the planet-generated spiral density
waves have quite weak signatures at near-infrared wavelengths
(Figure~\ref{fig:1um}).  This is because the outer edge of the gap
cleared by the planet casts a shadow on the outer disk, strongly
reducing the visibility of any spiral feature located outside the gap.
A similar result was obtained by \cite{Juhasz2015}, who constructed
synthetic scattered light images for a disk perturbed by planets,
assuming that the disk pressure scale height scales as a power law
with the radius (their Figure~2).  As shown in Figure~\ref{fig:1um},
the spirals' visibility at infrared wavelengths is even worse when the
pressure scale heights are consistent with hydrostatic equilibrium.

Another problem is the large pitch angles of the spiral arms observed
in the near-infrared.  Since the pitch angle of a spiral density wave
launched by a planet increases with the gas temperature (see
Section~\ref{sec:p-d}), large angles imply high temperatures.  By
comparing MWC~758 observations with the analytical solution of
Equation~\ref{eq:raf}, \cite{Benisty2015} inferred that the disk
temperature at the location of the spirals should be several hundred
degrees Kelvin, at least ten times greater than the temperature
calculated based on the stellar flux received.  How can the disk reach
such high temperatures?  \citet{Lyra2016} suggest the disk might be
heated as the density waves launched by a massive planet (5~M$_J$)
shock the circumstellar gas.  Note that this works only if the heat is
not radiated away, but remains in the disk from one shock passage to
the next -- that is, $t_{th}$ must exceed $t_{dy}$, which in our
initial disk model happens only within 10~AU of the star.

Furthermore, the shock heating is deposited preferentially near the
planet's innermost Lindblad resonances, which lie close to the edges
of the gap the planet opens in the disk \citep[Figure~3
  of][]{Lyra2016}.  Hot gas at the Lindblad resonances would mean a
taller outer edge for the gap, which could even reduce the visibility
of spiral waves in the scattered light emission.  It is also not yet
clear whether such hot material would produce hot spots visible in the
mid-infrared or millimeter continuum emission.  ALMA observations
constrain the temperatures of dust and gas (most gas measurements are
of carbon monoxide) near the spiral arms to less than 40~K in
agreement with predictions from stellar irradiated disk models
\citep[][]{2015PASJ...67..122M, 2017arXiv171208845B}.
 
As a solution to both the spirals' visibility and the pitch angle
problem, \cite{Dong2015b} have suggested the observed spiral waves
might be excited by a massive planet ($>5$M$_J$) orbiting outside the
cavities observed at millimeter wavelengths, and beyond the spiral
features observed in the near-infrared.  In this model, the observed
spirals correspond to the arms propagating inward from the planet
orbital radius.  This is similar to the picture proposed by
\cite{Muto2012} to explain the large pitch angle of the spiral
features observed in the SAO~206462 disk.  In the case of MWC~758,
\cite{Dong2015b} calculate that the planet would have an orbital
radius of about 160~AU (their Figure~4).  For this model to work, the
planet must be very young.  In particular, the age of the planet must
be less than the time required to open a gap in the disk.  This is
because once the gap is fully opened, the part of the spiral density
wave with the largest pitch angle will be located within the dust
depleted gap, making the spiral disappear at infrared wavelengths.  In
the specific case of MWC~758, \cite{Dong2015b} manage to reproduce the
observations by halting the hydrodynamic simulation after 20 orbits of
the planet, or about $0.3\times 10^5$~yr for an orbital radius of
160~AU.  The entire gap opening phase takes $0.6\times 10^5$ to
$1.5\times 10^5$~yr, depending on the accretion stress.  If the
observable lifetime of the spiral is so short, objects like MWC~758
and SAO~206462 should be a small fraction, probably less than 10\%, of
the disk population forming giant planets at large separation.  This
seems to conflict with the fact the spiral arcs are observed in the
majority of perturbed disks around early type stars.

A third hypothesis, advanced by \cite{Dong2015d}, is that the spirals
observed in scattered light result from the disk's gravitational
instability.  To be gravitationally unstable, disks must be massive.
\citet{Dong2015d} find that the spiral structures in SAO~206462 and
MWC~758 require disk masses of 0.25 and 0.5~M$_\odot$, respectively.
These are much greater than the masses derived from observations of
the millimeter-wave dust emission, which suggest instead masses of
about 0.025 and 0.01~M$_\odot$, respectively
\citep{Isella10b,Andrews11a}.

Finally, a fourth hypothesis is that the spiral features are due to
material orbiting near the star which casts shadows that appear as
spirals because they sweep across the disk at the finite speed of
light \citep{2016arXiv160803147K}.  Forming bright spiral arms this
way requires notches in the screening material to let through narrow
beams of starlight, while variability surveys suggest narrow-angle
obscuration is more common \citep{2014AJ....147...82C,
  2015AJ....149..130S}.

Clues to the causes of the spiral features observed in the
near-infrared might be gained by looking at the ensemble of perturbed
disks listed in Table~\ref{tab:disks}.  For example, as noted above,
five of the eight disks around A and F stars are characterized by
spiral features, while no spirals are detected in the disks of the K
type stars for which observations exist.  Whereas we cannot exclude
that this trend might be an artifact of the small sample, the
available observations suggest that the formation of spiral arms
somehow depends on the stellar mass.
  
It is worth noting that early-type stars are mostly in binary or
multiple systems.  The multiplicity fraction of mature G-type stars is
57\%, complete down to a companion-to-primary mass ratio of 0.1.  This
fraction is higher among A- and F-type stars.  The large fraction of
multiples among the main-sequence stars suggests a correspondingly
high fraction of multiples among young early-type stars.  Based on
this consideration, we suggest that the observed perturbations might
be due to the presence of yet unseen, close separation, stellar mass
companions, instead of planets.  A companion with greater mass will
excite stronger density waves, which might warm up the circumstellar
gas through shock heating, as remarked above, or induce other kinds of
instability in the disk.  The heating from the companions should also
be taken into account in future hydrodynamical modeling of this
scenario.

A companion might be responsible for at least one of the
dynamically-hot disks listed in Table~\ref{tab:disks}.  The star
HD~142527 has a 0.25~M$_\odot$ companion orbiting at 10~AU.  This
binary system has a mass ratio of 8, much lower than those in the
models we present here, which range from 330 to~16700.  Searches for
companions around the other dynamically-hot disks listed in
Table~\ref{tab:disks} have resulted in no detections to date, but most
of these observations are sensitive mostly to companions at
separations larger than 10~AU \citep[][and references
  therein]{2014ApJ...788..129I}.

\section{Conclusions}

Protostellar disks show central cavities, bright and dark rings, and
spiral arms that could be from gravitational perturbations by low-mass
stellar or planetary companions, which are expected to be nearly
ubiquitous.  However, many of the features can also be made by
processes intrinsic to the disks, with no companion bodies present.
Few or no planets have yet been conclusively identified embedded in
protostellar disks, and attempts to connect the observed spiral arms
to planets on particular orbits appear to require either rather high
disk temperatures, or that we have caught the planets in the
short-lived stage when they are crossing the threshold mass for
opening a gap.  Non-planet mechanisms also have difficulties.
Gravitational instability suffers from requiring disk masses much
larger than measured, while light-travel-time effects make bright
spirals only with specific patterns of obscuration by material near
the star.

To address these issues we have explored how young planets alter
protostellar disks' emission at near-infrared, millimeter, and
centimeter wavelengths, using 2-D vertically-averaged hydrodynamical
modeling to map the surface density around the planet, and 3-D
radiative transfer calculations to obtain the temperature structure,
from which we find the vertical distribution of material.  We consider
planets with masses just at the threshold for tidally clearing a gap
around their orbit, and planets with masses seven and seventy times
greater, which open substantial gaps.  We obtain model disks that are
in radiative equilibrium with the starlight heating, and in vertical
hydrostatic balance.  The planets modify the disks' structure and
appearance as follows:
\begin{enumerate}
\item A planet massive enough to open even a partial gap in the disk
  lets additional scattered starlight and re-emitted infrared
  radiation reach and warm the midplane.
\item The light scattered to our telescopes reveals the parts of the
  disk directly lit by the star.  The starlight's grazing angle of
  entry means that even small features on the disk surface cast long
  shadows.  The planet-carved gap's outer wall in particular receives
  extra starlight heating and puffs up, throwing a shadow across the
  disk beyond.  The shadow appears in scattered light as an additional
  dark ring, which could be mistaken for a gap opened by another more
  distant planet.
\item The shadow is darker and colder in models with the disk placed
  in vertical hydrostatic equilibrium, than in those where the scale
  heights are left unchanged from the disk without planets.  Our
  hydrostatic model with 100-M$_\oplus$ planet has a scattered light
  surface brightness contrast between gap outer wall and shadow that
  is about 5 times greater than the model where the temperatures and
  thus scale heights are as in the planet-free disk.  The contrast
  increases by an order of magnitude for the 1000-M$_\oplus$ planet
  model. 
\item For a disk mass and size rather typical of nearby protostellar
  disks, the millimeter emission arising from the regions where most
  planets are expected to form is optically thick, and therefore tells
  us about the dust temperature rather than the surface density.  The
  same disk is mostly optically thin at centimeter wavelengths, which
  therefore trace perturbations in the dust surface density such as
  those induced by planets.  Combining sensitive observations at
  millimeter and centimeter wavelengths is therefore key to measuring
  the dust temperature and density in the planet-forming regions of
  nearby disks \citep{2018arXiv180101223R}.
\item The shapes and contrast levels of the brightest areas in the
  synthetic millimeter and centimeter continuum images depend on
  whether we enforce hydrostatic equilibrium, even though this leaves
  surface densities unchanged, because the altered scale heights
  change the pattern of starlight illumination, causing shifts in
  temperature.
\end{enumerate}
A common theme in these results is that when interpreting protostellar
disk features as caused by an embedded planet, we cannot safely assume
the temperature at each position is fixed, but must consider radiative
heating and cooling.  The planet disturbs the disk, changing where the
starlight falls, which changes the temperatures, which further alters
the shape of the disk's surface.  This cascade of effects impacts the
architecture of the nascent planetary system: the temperature
gradients in the shadows cast by the puffed-up outer rims of the gaps
opened by our more massive planets are such that additional low-mass
planets' orbital migration will converge in the shadows.

The coupling of dynamics with radiative transfer which we have
explored will help in understanding the concentric rings observed in
some disks.  However several aspects of the observations remain
mysterious.  In particular, we have no good explanation for the spiral
features.  In fact, we have turned up evidence that they are probably
not caused by planets, since under hydrostatic equilibrium the outer
rim of the planet-opened gap casts a shadow so deep, it largely hides
the outer arm of the planet's spiral wave.  Also still unclear is why
more-massive stars often show spiral arms, while low-mass stars' disks
typically have azimuthally symmetric rings.

Two limitations of the models presented here are worth mentioning.
The first is connected with the millimeter and centimeter emission,
which come from sub-millimeter to millimeter-sized dust grains that
are dynamically well-coupled only to dense gas near the planet
(Section~\ref{sec:dust-gas}).  Since these big grains' interaction
with the gas depends on their poorly-known aerodynamic properties and
the distribution of turbulence, we have focused on the limit where the
dust and gas are well-mixed, leaving treatment of the gas-dust
dynamical interaction for future work.  The second limitation is that
radiative and hydrostatic equilibrium hold only in patches of disk
receiving time-steady illumination.  Any non-axisymmetric
disturbances' orbital motion sweeps their shadows across the material
beyond.  The illumination may vary faster than the outer disk can
respond, in which case some parts of the system will be perpetually
out of hydrostatic balance, always shrinking or expanding towards the
scale height consistent with their momentary temperature.  Capturing
this effect would require coupling the differential rotation with the
heating and cooling.

\acknowledgments A.I.\ and N.J.T.\ thank Mario Flock, Cornelis
Dullemond, Thomas Henning, Wilhelm Kley, Wladimir Lyra and Roy van
Boekel for helpful discussions, and Ruobing Dong, Nienke Van der Marel, and 
Myriam Benisty for sharing images  of SAO206463 and J1604. A.I. \ and N.J.T\
acknowledge support from the NASA
Origins of Solar Systems program through award NNX15AB06G and from the
JPL Research \& Technology Development Program through award
R.16.183.037.  A.I.\ acknowledges support from the NSF Grants
Nos.\ AST-1535809 and AST-1715718.  This work was carried out in part
at the Jet Propulsion Laboratory, California Institute of Technology,
under contract with NASA.

\appendix

\section{FARGO setup\label{appendix:fargo}}
The hydrodynamic calculations of the planet-disk interaction are
performed using the GPU version of the FARGO3D code.  Simulations are
run in two dimensions using a polar grid consisting of 1500~cells in
both the radial and azimuthal directions, for a total of $2.25\times
10^6$~cells.  The radial grid is linearly spaced and extends from 0.3
to 3~times the planet's orbital radius.  Each cell's radial size is
0.02~AU.  By comparison, the disk pressure scale height at the
position of the planet is 0.43~AU (Equation~\ref{eq:h_p}),
corresponding to 22~cells.

The disk feels the pull of the planet's gravity, while the planet is
fixed on its initial circular orbit at 10~AU.  The planetary
gravitational potential is smoothed by setting the {\it
  ThicknessSmoothing} parameter to~0.6.  Following
\cite{2012A&A...546A..99K}, this value of the potential smoothing
length provides the best agreement between two- and three-dimensional
hydrodynamical calculations.  The kinematic viscosity is set to zero
across the entire disk.  We begin each simulation by running for about
100~orbits while gradually increasing the planet's mass from zero to
its final value.  We then let the simulation run another 300~orbits,
by which time the surface density map is almost time-steady.

\section{RADMC-3D setup\label{appendix:radmc}}

The disk temperature and emitted radiation are calculated using the
Monte Carlo radiative transfer code RADMC-3D available at
\url{http://www.ita.uni-heidelberg.de/~dullemond/software/radmc-3d}.
We adopt spherical coordinates with 500~cells in the radial direction,
370~cells in the azimuthal direction, and 160~cells in the polar
direction.  The radial cells are equally spaced between 0.5 and 50~AU,
corresponding to a radial extent of about 0.1~AU.  The azimuthal grid
extends from 0 to 2$\pi$~radians, and the polar grid from~60\arcdeg\ to
120\arcdeg, where 0\arcdeg\ is the disk's rotational axis, and
90\arcdeg\ is the midplane.  We have performed several tests to make
sure the transfer calculations' results are unaffected by the number
of cells and the extent of the polar grid.

Each Monte Carlo simulation involves $10^9$~photon packets.  Computing
the radiative equilibrium temperature is by far the slowest part of a
time step in the poor man's radiation hydrodynamics scheme, requiring
about 1~hr of wall-clock time on a machine equipped with 20~Intel Xeon
2.8-GHz processors.  The midplane, which is the optically thickest
part of the disk, and so the least often visited by diffusing photon
packets, is then well enough sampled to reduce the temperatures'
statistical noise below 5\% everywhere.

\section{Poor Man's Hydrodynamics\label{appendix:pmhd}}

The novel feature of our simplified radiation hydrodynamics scheme is
the time-dependent approach to hydrostatic equilibrium, which we here
describe in detail.  Hydrostatic balance means the pressure gradient
matches gravity in the vertical direction,
\begin{equation}\label{eq:hse}
  {dp\over dz} = -\rho\Omega^2z.
\end{equation}
To solve eq.~\ref{eq:hse} for the new density profile $\rho^{n+1}(z)$,
we must specify how the pressure depends on the density.  Since the
pressure is proportional to density times temperature, we have to
guess the new temperature profile.  We need an estimate $T^*(z)$ that
is close to the next time step's radiative transfer solution,
$T^{n+1}(z)$.  A reasonable guess is that the $(n+1)$-th time step's
temperature will be the same as at the corresponding mass column in
the $n$-th time step, if the column is proportional to optical depth
and the starlight sets the temperature.  We define the column
\begin{equation}\label{eq:column}
  m(z)\equiv\int_z^\infty\rho(z) dz
\end{equation}
and build a lookup table recording how in the $n$-th time step
$m^n(z)$ maps to $T^n(z)$.  Then, we use the table to guess $T^*(z) =
T^n(m^{n+1}(z))$.

The rest of the procedure follows from discretizing eq.~\ref{eq:hse}
along the $z$-direction.  Let the spatial index $k$ run from zero in
the cell adjacent to the midplane, to $K$ in the cell just below and
touching the top boundary.  Then
\begin{equation}\label{eq:difference}
  {p_{k+1}-p_k\over \Delta z} =
  -\frac{1}{2}(\rho^{n+1}_{k+1}+\rho^{n+1}_k)\Omega^2z_{k+\frac{1}{2}}.
\end{equation}
where we use the approximation $\Omega^2\approx GM/R^3$, valid when
the disk is geometrically thin.  The pressure at the center of the
$k$-th cell $p_{k} = {\cal R}T^*_k\rho^{n+1}_k/\mu$, where as usual
${\cal R}$ is the gas constant and $\mu$ the mean molecular weight.
We solve eq.~\ref{eq:difference} for $\rho^{n+1}_{k+1}$ by specifying
$\rho^{n+1}_k$, $T^*_k$ and how $p_{k+1}$ depends on the choice of
$\rho^{n+1}_{k+1}$ through the temperature $T^*_{k+1}$, following
these steps:
\begin{enumerate}
  \item Start by guessing the density $\rho^{n+1}_{k=0}$ in the cell
    touching the midplane.
  \item Find the temperature $T^*_k$ using the lookup table.  To get
    the mass column $m^{n+1}_k$, recall that the surface density is
    fixed, so $m(z) = \frac{1}{2}\sigma - \int_0^z\rho(z) dz$, or in
    discretized form,
    \begin{equation}
      m_k=\frac{1}{2}\sigma
      - \sum_{k^\prime=0}^{k-1}\rho^{n+1}_{k^\prime}\Delta z
      - \frac{1}{2}\rho^{n+1}_k\Delta z,
    \end{equation}
    where the last term places us at the midpoint of cell $k$.
  \item Initially guess that the next cell's temperature $T^*_{k+1}$
    equals the lookup table value for the column $m_{k+\frac{1}{2}}$
    found at the top of cell $k$.
  \item Obtain the corresponding density by solving
    eq.~\ref{eq:difference} for $\rho^{n+1}_{k+1}$.  Now we can
    estimate the column at the center of cell $k+1$, and so from the
    lookup table a revised temperature $T^*_{k+1}$.
  \item Repeat step~4 till $T^*_{k+1}$ stops changing.  We've found a
    mutually consistent temperature and density for cell $k+1$.
  \item Move up one level to cell $k+2$, and carry out the same
    procedure from step~3.
  \item On reaching the upper boundary $k=K$, check whether the
    density profile has the required surface density $\sigma$.  If
    not, guess a new midplane starting density, zeroing in on the
    required value through bisection, and return to step~1.
\end{enumerate}
Note this procedure involves nested loops.  The outer one (steps 1--7)
finds the midplane density, and the inner one (steps 3--5) finds the
temperature in the next cell.

To implement poor man's radiation hydrodynamics, in step~5 we also
compare the density scale height at the cell boundary,
\begin{equation}\label{eq:H}
  {\hat H}^{n+1}_{k+\frac{1}{2}} \equiv
    {\frac{1}{2} (\rho^{n+1}_k + \rho^{n+1}_{k+1}) \Delta z
    \over \rho^{n+1}_k - \rho^{n+1}_{k+1}},
\end{equation}
with that at the same point in the previous time step $n$.  We prevent
the atmosphere from expanding or contracting from one time step to the
next any faster than either the sound-crossing or the heat diffusion
timescale, by limiting the new density scale height using
\begin{equation}\label{eq:deltaH}
  H^{n+1}_{k+\frac{1}{2}} - H^n_{k+\frac{1}{2}} =
  \left({\hat H}^{n+1}_{k+\frac{1}{2}} - H^n_{k+\frac{1}{2}}\right) \times
  \min\left(1, {0.1 t^0_{cr} \over \max[t_{dy}, t^n_{th}]}\right).
\end{equation}
We insert the revised scale height in eq.~\ref{eq:H}, and solve for
the revised $\rho^{n+1}_{k+1}$.  Otherwise, we follow step~4 in taking
the temperature from the new column.  For simplicity, we use the
midplane thermal timescale $t^n_{th,0}$ at all heights.  We have also
experimented with increasing the Courant-like time step factor in
eq.~\ref{eq:deltaH} from 0.1 to 0.3 in a calculation with the
Saturn-mass planet.  The disk evolves very similarly, and reaches an
almost identical equilibrium, but of course the calculation is quicker
because fewer time steps are needed.  We use a time step factor of 0.1
in all other calculations.

\bibliographystyle{apj}
\bibliography{ref}

\end{document}